\documentclass{article}

\usepackage{microtype}
\usepackage{graphicx}
\usepackage{subcaption}
\usepackage{booktabs} 

\usepackage{hyperref}
\usepackage{enumitem}
\usepackage{multicol,multirow}
\usepackage{xcolor}
\newcommand{\bmcell}[2]{#1{\tiny$\pm$#2}\%}

\usepackage[preprint]{icml2026}

\usepackage{amsmath}
\usepackage{amssymb}
\usepackage{mathtools}
\usepackage{amsthm}

\usepackage{algorithm}
\usepackage{algpseudocode}
\usepackage{bm}

\usepackage[capitalize,noabbrev]{cleveref}

\theoremstyle{plain}

\theoremstyle{definition}

\theoremstyle{remark}

\usepackage[textsize=tiny]{todonotes}

\icmltitlerunning{ConforNets: Latents-Based Conformational Control in OpenFold3}
\usepackage[T1]{fontenc}
\usepackage[scaled=0.85]{beramono}
\begin{document}

\twocolumn[
  \icmltitle{ConforNets: Latents-Based Conformational Control in OpenFold3}



  \icmlsetsymbol{equal}{*}

  \begin{icmlauthorlist}
    \icmlauthor{Minji Lee}{cs}
    \icmlauthor{Colin Kalicki}{bio}
    \icmlauthor{Minkyu Jeon}{princeton}
    \icmlauthor{Aymen Qabel}{cs}
    \icmlauthor{Alisia Fadini}{bio}
    \icmlauthor{Mohammed AlQuraishi}{cs,bio}
  \end{icmlauthorlist}

  \icmlaffiliation{bio}{Department of Systems Biology, Columbia University, NY, USA}
  \icmlaffiliation{cs}{Department of Computer Science, Columbia University, NY, USA}
  \icmlaffiliation{princeton}{Department of Computer Science, Princeton University, Princeton, NJ, USA}

  \icmlcorrespondingauthor{Alisia Fadini}{af3659@cumc.columbia.edu}
  \icmlcorrespondingauthor{Mohammed AlQuraishi}{m.alquraishi@columbia.edu}

  \icmlkeywords{protein conformation prediction, diffusion models, alphafold3}

  \vskip 0.3in
]



\printAffiliationsAndNotice{}  

\begin{abstract}
Models from the AlphaFold (AF) family reliably predict one dominant conformation for most well-ordered proteins but struggle to capture biologically relevant alternate states. Several efforts have focused on eliciting greater conformational variability through \textit{ad hoc} inference-time perturbations of AF models or their inputs. Despite their progress, these approaches remain inefficient and fail to consistently recover major conformational modes. Here, we investigate both the optimal location and manner-of-operation for perturbing latent representations in the AF3 architecture. We distill our findings in ConforNets: channel-wise affine transforms of the pre-Pairformer pair latents. Unlike previous methods, ConforNets globally modulate AF3 representations, making them reusable across proteins. On unsupervised generation of alternate states, ConforNets achieve state-of-the-art success rates on all existing multi-state benchmarks. On the novel supervised task of conformational transfer, ConforNets trained on one source protein can induce a conserved conformational change across a protein family. Collectively, these results introduce a mechanism for conformational control in AF3-based models.
\end{abstract}

\section{Introduction}

The AlphaFold (AF) \citep{senior2020improved,jumper2021highly,abramson2024accurate} series of models has transformed protein biology by yielding accurate predictions of the native structure of most proteins. These models continue to, however, struggle to capture the conformational heterogeneity and context-dependent changes that underlie many protein functions. Two broad and highly active lines of research are currently attempting to tackle this limitation. The first focuses on learning the free energy landscapes of proteins by training models on experimental datasets of observed protein conformations augmented by synthetic datasets of molecular dynamics (MD) trajectories \citep{alphaflow,bioemu,roney2025protein}. While these models borrow, often heavily, architectural elements from AF2/3, they are generally trained from scratch and represent distinct models aimed at producing calibrated, Boltzmann-weighted ensembles. In contrast, the second line of research focuses on predicting the major states of a protein in an uncalibrated fashion, to provide useful hypotheses for downstream applications. Its methods have largely focused on inference-time perturbations of AF models with no or minimal training.

While the first approach is broader in scope and more physically grounded, it is constrained by the limited lengths of available MD trajectories, which typically do not sample slow but biologically relevant conformational changes. The resulting learned energy landscapes thus often miss major states that the second category of methods pragmatically attempt to discover. This latter, "perturbational" category, is underpinned by a hypothesis---one we share---that AF models internalize salient aspects of a protein's conformational landscape despite being trained to only predict its native state. \textls[-10]{We believe this occurs because AF2/3 encounter during training alternate states of the same protein, either explicitly tied to its sequence or manifesting as the native states of homologous proteins whose energy landscapes have sufficiently shifted to prefer the alternate states. This is reinforced by AF's training regimen, which subsamples input multiple sequence alignments (MSAs), preventing these models from relying exclusively on the MSA's co-variation signal, and thus encouraging them to learn an implicit physical function that maps protein sequences to their conformational preferences.}

Perturbational methods elicit alternate states by operating on the inputs, latents, and structure generation processes of AF2 and AF3. Empirical results increasingly suggest that such approaches can be effective \citep{del2022sampling,kalakoti2025afsample2,wayment2024predicting,richman2025unlocking}. A key determinant of performance is where and how perturbations are applied. Existing approaches seldom operate optimally, leading to inefficient sampling, missing conformational modes, and implausible structures with physical violations. Most efforts have focused on perturbing input MSAs  \citep{wayment2024predicting,lee2025large,kalakoti2025afsample2}, owing to how strongly evolutionary couplings inform protein structure. However, the space of these perturbations is combinatorially vast and constrained by the depth and diversity of the starting MSA. Perturbing latents, where implicit conformational knowledge may be encoded, is natural but not guaranteed to yield energetically accessible and physically plausible states. The same is arguably truer when directly perturbing predictions in coordinate space, for instance in AF3’s diffusion module \citep{richman2025unlocking}, which can be brittle \citep{li2026robust} due to bypassing the extensive structural reasoning performed prior to diffusion.

In this work, we introduce ConforNets, a new inference-time, perturbational approach that optimizes the placement and method-of-operation on the AF3 architecture. Through thorough experimentation, we show that the latent pair representation preceding the Pairformer offers the greatest control (Sec.~\ref{sec:results_perturbation}), likely by enabling the Pairformer to alter its processing of residue-residue contacts (App.~\ref{app:foldswitch_interp}). Unlike existing perturbational methods, which operate residue-wise and are thus protein-specific, ConforNets are channel-wise affine transforms. They modulate AF3’s latents globally and permit reuse across different proteins. ConforNets are broadly applicable to all AF3-based models, and we use OpenFold3-preview (\textbf{OF3p}) \citep{of3p}, an open-source reproduction of AlphaFold3 and its associated pretrained weights, as our implementation chassis. We develop two types of ConforNets to tackle two problems:

\begin{enumerate}

\item \textbf{Unsupervised prediction of alternate conformations}. We jointly optimize multiple ConforNets to maximize pairwise distances between their generated samples, inducing them to collectively learn to sample distinct modes of the conformational landscape. On multi-state benchmarks covering 104 pairs of diverse conformational changes, ConforNets outperform all other methods (Sec.~\ref{sec:diversity_results}). ConforNets optimization is fast and easy to integrate into OF3p inference, requiring less than 40 GPU seconds for a 200-residue protein.

\item \textbf{Supervised transfer of conformational state}. Many proteins undergo conformational changes that are conserved across their families, such as activation of GPCRs or the DFG-in/out transition of kinases, both common drug targets. At-will induction of these states can facilitate downstream applications such as docking and mutational analysis, and may even provide mechanistic structural hypotheses. Standard AF3 inference typically produces a single state, and, when it does not, provides no control over what state is produced (Fig.~\ref{fig:problem_intro}). We introduce the concept of supervised transfer of conformational state, where we train a ConforNet on a (single) source protein that encodes a specific state, then use this ConforNet to induce the same state in other proteins of the same structural family. When tested on three protein families, ConforNets substantially outperform standard inference on this novel task (24$\rightarrow$79\% for GPCR active, 6$\rightarrow$23\% for kinases DFG-out, and 16$\rightarrow$57\% for transporters outward-open).
\end{enumerate}

\section{Related Work}
\label{sec:related_works}
\subsection{Prediction of alternate protein conformations}
Efforts aimed at producing diverse protein states fall into four categories. We review and relate them to ConforNets.

\paragraph{(i) Calibrated ensembles by explicit training}
One line of work adapts structure prediction methods to produce Boltzmann-weighted conformational ensembles by training on MD trajectories in combination with datasets of experimentally observed alternate states. AlphaFlow \citep{alphaflow} repurposes the AF2 trunk (using OpenFold \citep{ahdritz2024openfold} or ESMFold \citep{lin2023evolutionary}) as a sequence-conditioned flow-matching model, and trains the entire system end-to-end to predict denoised protein backbone coordinates from the PDB. BioEmu \citep{bioemu} trains a backbone diffusion model conditioned on fixed AF2 Evoformer embeddings, using sequence-to-ensemble training pairs derived from AFDB \citep{varadi2022alphafold}. By virtue of requiring explicit ensembles as training data, these models are highly dependent on the quality and quantity of said data, which remains exceedingly rare for experimental sources, and short in timescales for MD trajectories.

\paragraph{(ii) Uncalibrated ensembles by implicit learning}
A second line of work seeks to generate conformational ensembles (not necessarily Boltzmann-weighted) without explicit ensemble supervision by implicitly capturing conformational variability through specialized training procedures. ESMDiff \citep{lu2024structure} learns a language model that maps amino acid sequences to learned structure tokens whose latent representations inherently encode conformational variability. ConfDiff \citep{wang2024protein} incorporates guidance from an MD force field into an SE(3)-equivariant diffusion model conditioned on ESMFold embeddings derived exclusively from PDB structures. In both cases, the models capture variation that is limited in spatial scale, as they are not trained, tuned, or steered to recover large transitions. We consider AF2/3 training to fall into this category.

\paragraph{(iii) MSA perturbation}
In AF2/3-derived models, conformational variability can often be elicited by subsampling input MSAs during inference, presumably due to the altered evolutionary couplings encoded in the subsampled MSA. Building on this observation \citep{del2022sampling}, several methods perturb input MSAs to increase conformational diversity. \citet{stein2022speach_af} mutate high-confidence interacting positions in the MSA, inducing the model to attend to alternate couplings. AFcluster \citep{wayment2024predicting} observes that distinct phylogenetic branches may be biased toward different states, and effectuates this observation by clustering the MSA before performing inference using single clusters that isolate branch-specific evolutionary couplings. CF-random \citep{lee2025large} counters by asserting that state diversity can be elicited through shallow MSA subsampling alone, without explicit clustering \citep{schafer2025sequence}. AFsample2 \citep{kalakoti2025afsample2} introduces random MSA column masking and dropout, and AFsample3 \citep{kalakoti2026afsample3} extends this to AF3. While they provide partially effective heuristics, these methods must contend with a vast combinatorial space of MSA perturbations---without a mechanism to specify the type or magnitude of induced conformational change---and with their effectiveness being contingent on the depth and diversity of the starting MSA.

\paragraph{(iv) Diffusion guidance at inference time}
A fourth line of work focuses on inference-time modulation of AF3's diffusion module. ConforMix \citep{richman2025unlocking} uses diffusion guidance to bias AF3 towards states with fixed target RMSDs to the baseline AF3 prediction. ConforMix and our work share a common motivation of using intentional objectives to steer AF3 sampling in lieu of uncontrolled random input perturbations. They differ in how and where this control is applied. ConforMix operates on AF3's diffusion module while ConforNets intervene upstream, in the Pairformer, via channel-wise transforms. We argue that operating in this way is appealing for two reasons. First, ConforNets alter the conditioning signal, not the score function, allowing AF3's diffusion module to map latents onto the geometrically valid protein structure manifold as it normally does. Second, because ConforNets are learned transforms, once trained they are reusable across proteins in a way that is inherently inaccessible to diffusion guidance.

\begin{figure*}[t!]
    \centering
    \includegraphics[width=\linewidth]{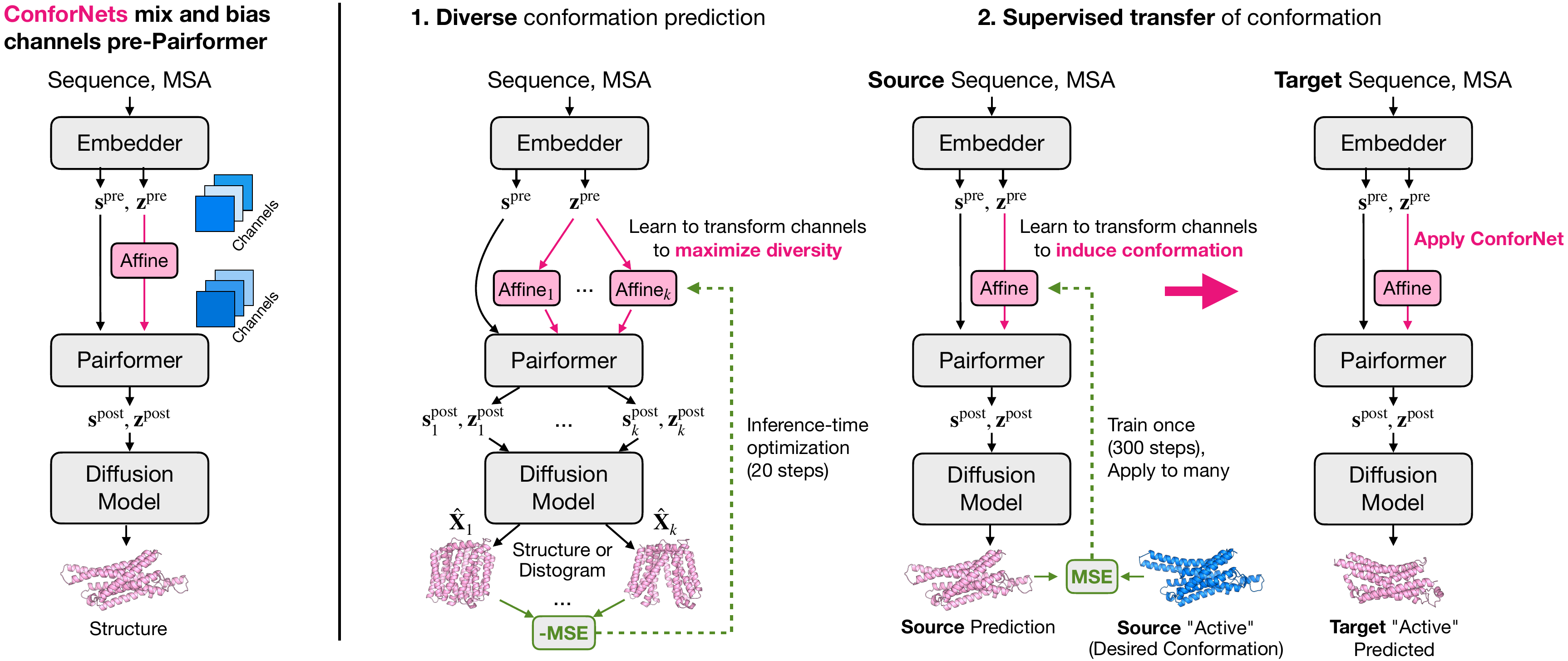}
    \caption{\textbf{ConforNets induce conformations by mixing and biasing the channels of AF3's pair latents to satisfy different objectives}. \textbf{1. Diverse conformation prediction}: $k$ ConforNets are jointly trained to transform $\mathbf{z}^{\text{pre}}$ such that the diversity of predicted conformations is maximized. Optimization is done via gradient descent using distogram- or coordinates-based objectives. \textbf{2. Supervised transfer of conformation}: one ConforNet is trained to minimize the MSE between a desired conformation and the predicted structure of a source protein. The trained ConforNet is then applied to other proteins (of varying lengths) to induce a similar conformational state.}
    \label{fig:method}
\end{figure*}

\subsection{Latents optimization in AF series models}
Several methods optimize and perturb AF2/3 latents to steer structure prediction towards a desired goal. \citet{wu2025robust} perform a single gradient update on AF2's MSA and pair latents to maximize distogram entropy, inducing alternate states in membrane proteins. \citet{bryant2024improved} optimize MSA profiles in AF-Multimer \cite{evans2021protein} to maximize the model's predicted confidence (pLDDT), yielding improved multimeric complexes. \citet{fadini2026alphafold} optimize AF2's MSA profiles, while \citet{li2026robust} and \citet{maddipatla2026inference} update AF3's Pairformer latents to improve agreement between predicted structures and experimental observables, injecting priors from crystallographic, cryo-EM/ET, and NMR datasets. All of the above approaches directly optimize AF2/3 representations (across the sequence dimension) during inference and are therefore inherently per-protein procedures.

\subsection{Relation to methods in other ML domains}
ConforNets are conceptually related to task/capability-specific adaptation methods such as low-rank adaptation (LoRA) \citep{hu2022lora}, prompt tuning \citep{lester2021power,li2021prefix}, and ControlNet \citep{zhang2023adding}. While widely explored in domains such as language and image generation, these strategies have seen limited use in the molecular sciences. Our work suggests that pretrained protein structure predictors implicitly encode rich physical priors, permitting lightweight adaptations that unlock new generative capabilities without additional training.


\section{Method}
\label{sec:method}

We introduce channel-wise affine transformations that modulate OF3p latent representations to shape its conformational preferences. We find that transforming the pair representation preceding the Pairformer trunk delivers the greatest degree of conformational control.

\subsection{OF3p inference and diffusion}
\label{sec:af3}

In standard OF3p inference, a protein sequence $x$ of length $L$ and associated MSA $\mathcal{M}$ are initially embedded to single and pair representations:
\begin{equation*}
    \mathbf{s}^{\mathrm{pre}},\mathbf{z}^{\mathrm{pre}}=\texttt{Embedder}(x,\texttt{Subsample}(\mathcal{M})),
\end{equation*}
where $\mathbf{s}^{\mathrm{pre}} \in \mathbb{R}^{L \times c_s},
\mathbf{z}^{\mathrm{pre}} \in \mathbb{R}^{L \times L \times c_z}$. \texttt{Subsample} denotes random subsampling to 1,024 MSA rows, unless otherwise specified. The Pairformer then refines the two representations via triangular updates and self-attention, yielding 
\begin{equation}
\label{eq:abstraction_pairformer}
    \mathbf{s}^{\mathrm{post}}, \mathbf{z}^{\mathrm{post}} \;=\; \texttt{Pairformer}\!\left(\mathbf{s}^{\mathrm{pre}}, \mathbf{z}^{\mathrm{pre}}\right).
\end{equation} 
The diffusion model then predicts structures conditioned on $\mathbf{s}^{\mathrm{pre}}$, $\mathbf{s}^{\mathrm{post}}$, and $\mathbf{z}^{\mathrm{post}}$, which are provided to the denoiser at every diffusion step together with the current noisy coordinates. OF3p adopts the Elucidated Diffusion Model (EDM) parameterization \citep{karras2022elucidating}, which defines the destructive process as continuous-time variance-exploding diffusion. The denoiser predicts a clean structure from a noisy input at a given $\sigma$, enabling both stochastic solvers and deterministic solvers. As in standard OF3p inference, we refer to sampling using an Euler-Maruyama-like solver with 200 discretizations as a full rollout. For monomeric proteins, however, we observe that a single denoising step often produces a reasonable structure, which we exploit for efficient optimization. From $\mathbf{z}^{\mathrm{post}}$, OF3p also predicts distogram logits $\mathbf{d} \in \mathbb{R}^{L \times L \times D}$ that parameterize categorical distributions over inter-residue distances.

\subsection{ConforNet}

We formulate a ConforNet $\phi$ that adapts a latent $\mathbf{h}$ with channel dimension $c$ (\textit{e.g.},\ $c_z = 128$ for pair latents) as an affine transform $
\phi(\mathbf{h})
\;=\;
\mathbf{h}\mathbf{W}^\top + \mathbf{b},
$ with trainable parameters $\mathbf{W} \in \mathbb{R}^{c \times c}$ and $\mathbf{b} \in \mathbb{R}^{c}$. ConforNets are initialized to the identity ($\mathbf{W} = \mathbf{I}$, $\mathbf{b} = \mathbf{0}$).

\subsubsection{Diverse conformation prediction}
\label{sec:training}

Given a single input protein, we jointly optimize $k$ ConforNets $\{\phi_1, \dots, \phi_k\}$ to produce $k$ maximally distinct structures. We add Gaussian noise to ConforNets after initialization to break symmetry. ConforNets transform the same pair latents before they are passed through the Pairformer:
\begin{equation*}    
(\mathbf{s}^{\mathrm{post}}_i, \mathbf{z}^{\mathrm{post}}_i)
\;=\;
\texttt{Pairformer}\!\left(\mathbf{s}^{\mathrm{pre}},\, \phi_i(\mathbf{z}^{\mathrm{pre}})\right).
\end{equation*}

From each $\mathbf{z}^{\mathrm{post}}_i$, OF3p predicts distogram logits $\mathbf{d}_i$. For objectives defined on coordinates, we use a single deterministic denoising step to predict atom coordinates.

\paragraph{Optimization objectives.}
We incentivize diversity in ConforNet-induced structures by maximizing distance in distogram or coordinate space. We also consider two baselines: maximizing distance in pair representation space and maximizing distogram entropy \citep{wu2025robust}.

\begin{enumerate}[noitemsep, topsep=0pt]
    \item \textbf{Distogram CDF MSE}. From distogram logits $\mathbf{d}_i$, we compute probabilities $\mathbf{p}_i = \mathrm{softmax}(\mathbf{d}_i)$ 
    and cumulative distributions $\mathrm{CDF}^{(i)} = \mathrm{cumsum}(\mathbf{p}_i)$.
    We then maximize the pairwise error between $\mathrm{CDFs}$:
    \[
    \mathcal{L}_{\mathrm{dist}}
    \;=\;
    -\sum_{i \neq j}
    \texttt{MSE}\!\left(
    \mathrm{CDF}^{(i)},\,
    \mathrm{CDF}^{(j)}
    \right).
    \]
    \item \textbf{Coordinate MSE}. We obtain one-step denoised structures $\widehat{\mathbf{X}}^{(i)}$. After rigid alignment, we maximize deviation in coordinate space:
    \[
    \mathcal{L}_{\mathrm{coord}}
    \;=\;
    -\sum_{i \neq j}
    \texttt{MSE}\!\left(
    \texttt{Align}(\widehat{\mathbf{X}}^{(i)}),\,
    \widehat{\mathbf{X}}^{(j)}
    \right).
    \]
    \item \textbf{Pair representation MSE} (baseline).
    We encourage differences in latents by maximizing MSE between post-Pairformer pair representations:
    \[
    \mathcal{L}_{\mathrm{pair}}
    \;=\;-
    \sum_{i \neq j}
    \texttt{MSE}\!\left(
    \mathbf{z}^{\mathrm{post}}_i,\,
    \mathbf{z}^{\mathrm{post}}_j
    \right).
    \]
    \item \textbf{Entropy maximization} (baseline). Following \citet{wu2025robust}, we maximize the entropy of the predicted distogram (not a pairwise objective):
    \[
    \mathcal{L}_{\mathrm{ent}}
    \;=\;
    -\mathbb{E}_{i}\,\mathbb{E}_{(u,v)}
    \left[
    -\sum_{d=1}^{D}
    p^{(i)}_{uv}(d)\log p^{(i)}_{uv}(d)
    \right].
    \]
\end{enumerate}

\paragraph{Optimization details.}
We train $k=2$\footnote{We focus on $k=2$ because all benchmarks considered in this work contain only two reference conformational states.} ConforNets for 20 steps using the Adam optimizer \citep{kingma2014adam} with an initial learning rate of $0.001$, which we halve every 5 steps. We clip gradients to a norm of 10 per ConforNet. At each step, $\mathbf{s}^{\text{pre}}$ and $\mathbf{z}^{\text{pre}}$ are recomputed from different MSA subsamples, preserving the stochasticity of OF3p inference and promoting robustness in the learned ConforNets.

\subsubsection{Conformation transfer}
\label{sec:transfer}

Given a source protein $x$ with a desired reference conformation $\mathbf{X}_\text{ref}$, we optimize a ConforNet $\phi$ so that the one-step deterministic diffusion sample $\widehat{\mathbf{X}}_{\phi}$ from adapted representation $\phi(\mathbf{z}^{\mathrm{pre}})$ reconstructs the reference by minimizing
\begin{align*}
\mathcal{L}_\text{transfer}&=
\texttt{MSE}\!\left(
\texttt{Align}\!\left(\widehat{\mathbf{X}}_{ \phi}\right),
\mathbf{X}_\text{ref}
\right)\\
\text{where } &\mathbf{s}^{\mathrm{pre}},\mathbf{z}^{\mathrm{pre}}=\texttt{Embedder}(x,\texttt{Subsample}(\mathcal{M})), \\ &\widehat{\mathbf{X}}_{\phi}=\texttt{Diffusion}(\texttt{Pairformer}(\mathbf{s}^{\mathrm{pre}},\phi(\mathbf{z}^{\mathrm{pre}}))).
\end{align*}

$\mathbf{s}^{\mathrm{pre}}$ and $\mathbf{z}^{\mathrm{pre}}$ are recomputed from different MSA subsamples at each optimization step, as in the diverse state prediction setting above.
The learned ConforNet $\phi$ encodes a state bias towards $\mathbf{X}_\text{ref}$ in the Pairformer latents. To effect conformational transfer at inference time, we apply $\phi$ to the pre-Pairformer pair latents of any target protein (typically of the same family as $x$), propagate the transformed latents through the Pairformer, then perform full diffusion rollout.

\subsubsection{Notes on recycling}

OF3p refines single and pair representations through multiple recycles of Pairformer passes, typically 10. In our experiments, we apply ConforNets only in the final pass, consistent with OF3p training where gradients are enabled only in the last pass. For simplicity, Eq.~\ref{eq:abstraction_pairformer} abstracts this by using $\mathrm{pre}$ and $\mathrm{post}$ to denote the single and pair representations immediately before and after the final pass.

\section{Unsupervised Prediction of Conformations}
\label{sec:diversity_results}

\begin{table*}[htbp]
\small
\centering
  \caption{\textbf{Multi-state benchmarks.} Means and standard deviations of success@100 (using 100 bootstrap trials) for ConforNets and other methods. Ours-coord and Ours-dist denote ConforNets trained with coordinate MSE and distogram CDF MSE objectives, respectively.}
  \label{tab:main}
  \begin{tabular}{lcccccccc}
\toprule
Benchmark & $N$ & BioEmu & OF3p & Shallow & AFsample3 & ConforMix & Ours-coord & Ours-dist \\
\midrule
Cryptic pockets (apo) & 34 & \bmcell{30.5}{2.8} & \bmcell{30.8}{1.7} & \bmcell{34.6}{2.8} & \bmcell{44.7}{1.7} & \bmcell{37.0}{1.5} & \bmcell{48.2}{1.8} & \bmcell{\textbf{48.8}}{2.9} \\
Cryptic pockets (holo) & 34 & \bmcell{60.6}{3.2} & \bmcell{63.7}{2.6} & \bmcell{53.4}{2.6} & \bmcell{73.6}{2.3} & \bmcell{62.7}{2.0} & \bmcell{\textbf{83.0}}{2.5} & \bmcell{78.9}{2.8} \\
Domain motions & 42 & \bmcell{73.1}{3.1} & \bmcell{69.5}{2.5} & \bmcell{72.8}{2.6} & \bmcell{80.6}{2.4} & \bmcell{80.3}{1.6} & \bmcell{81.7}{1.3} & \bmcell{\textbf{81.9}}{1.4} \\
OOD60 & 38 & \bmcell{43.1}{2.5} & \bmcell{45.3}{1.9} & \bmcell{44.9}{3.6} & \bmcell{54.0}{3.0} & \bmcell{57.7}{3.3} & \bmcell{53.7}{3.2} & \bmcell{\textbf{60.7}}{2.8} \\
Membrane transporters & 30 & \bmcell{34.3}{3.2} & \bmcell{24.3}{2.1} & \bmcell{28.7}{2.8} & \bmcell{46.9}{3.4} & \bmcell{34.9}{2.5} & \bmcell{47.2}{3.3} & \bmcell{\textbf{51.1}}{3.7} \\
Fold switchers & 30 & \bmcell{43.1}{0.8} & \bmcell{52.7}{2.6} & \bmcell{47.3}{2.6} & \bmcell{48.7}{1.8} & \bmcell{\textbf{54.3}}{2.6} & \bmcell{52.5}{2.2} & \bmcell{\textbf{54.4}}{2.4} \\
\bottomrule
\end{tabular}
\end{table*}

\subsection{Benchmarks}
\label{sec:benchmarks}

In this section, we evaluate ConforNets' ability to predict, unsupervised, alternate protein conformations. We use a broad suite of multi-state benchmarks employed in BioEmu \citep{bioemu} and ConforMix \citep{richman2025unlocking}, comprising 104 proteins with two distinct and experimentally determined conformations (208 structures in total). Following prior work, we group proteins as follows:

\begin{itemize}[noitemsep, topsep=0pt]
  \item \textbf{Domain motions} (N=21): large-scale hinge motions. 
  \item \textbf{Membrane transporters} (N=15): inward-open vs.\ outward-open conformations; curated by \citet{kalakoti2025afsample2}.
  \item \textbf{Cryptic pockets} (N=34): apo vs.\ holo pockets; curated in BioEmu from datasets by \citet{cimermancic2016cryptosite} and \citet{meller2023accelerating}.
  \item \textbf{Fold switchers} (N=15): proteins adopting two distinct folds; curated by \citet{porter2018extant}.
  \item \textbf{OOD60} (N=19): proteins deposited after AF2 training date cutoff and out-of-distribution for BioEmu (but not necessarily for other methods); we treat this as a general-purpose benchmark.
\end{itemize}

\subsection{Experimental design}
\label{sec:baselines}

We (re)implement shallow MSA subsampling \citep{lee2025large}, ConforMix \citep{richman2025unlocking}, AFsample3 \citep{kalakoti2026afsample3}, and entropy guidance \citep{wu2025robust} under a common OF3p framework for fair comparison. We provide as input MSAs generated from the ColabFold \citep{mirdita2022colabfold} server without templates or ligands (App.~\ref{app:input_query}). Following standard OF3p inference, we subsample MSAs to a maximum of 1,024 rows per recycle, except for shallow MSA subsampling, when we subsample to 8 rows. For entropy guidance \citep{wu2025robust}, which in its original AF2-based form performs a single-step update to the MSA and pair latents, we observe limited effects under OF3p diffusion sampling. We therefore reformulate it as an entropy objective (Sec.~\ref{sec:training}) within the ConforNet framework. We use BioEmu checkpoint v1.2, which is fine-tuned on MD simulations and a folding free energy dataset \citep{tsuboyama2023mega}. See App.~\ref{app:baselines} for further details.

\paragraph{Seed count and diffusion rollouts} To preserve the conformational diversity generated by MSA subsampling, we use 160 random seeds and 5 diffusion rollouts for OF3p, shallow MSA subsampling, and AFsample3 (800 samples total). For ConforNets, we retain the 5 diffusion rollouts but use 20 seeds with $k=2$ and take structures after 5, 10, 15, and 20 ConforNet optimization steps to generate $20\times2\times4\times5=800$ samples. As benchmark targets span a wide range of conformational changes and we do not know \textit{a priori} at which step the desired conformation emerges, we sweep a range of steps to maximize diversity.

\paragraph{Pairformer passes $R$} We consider two recycling settings: $R=11$ passes (10 recycles) and $R=1$ passes (no recycles). In Table~\ref{tab:main}, we report the better of $R=11$ or $R=1$ for all methods, as some tasks benefit from fewer recycles (we consider this part of the fixed sampling budget). In all figures, unless otherwise noted, we use $R=11$. Comparisons between fixed $R=11$ and $R=1$ are provided in App~\ref{app:fixed_passes}.

\begin{figure}[t!]
    \centering
    \includegraphics[width=.96\linewidth]{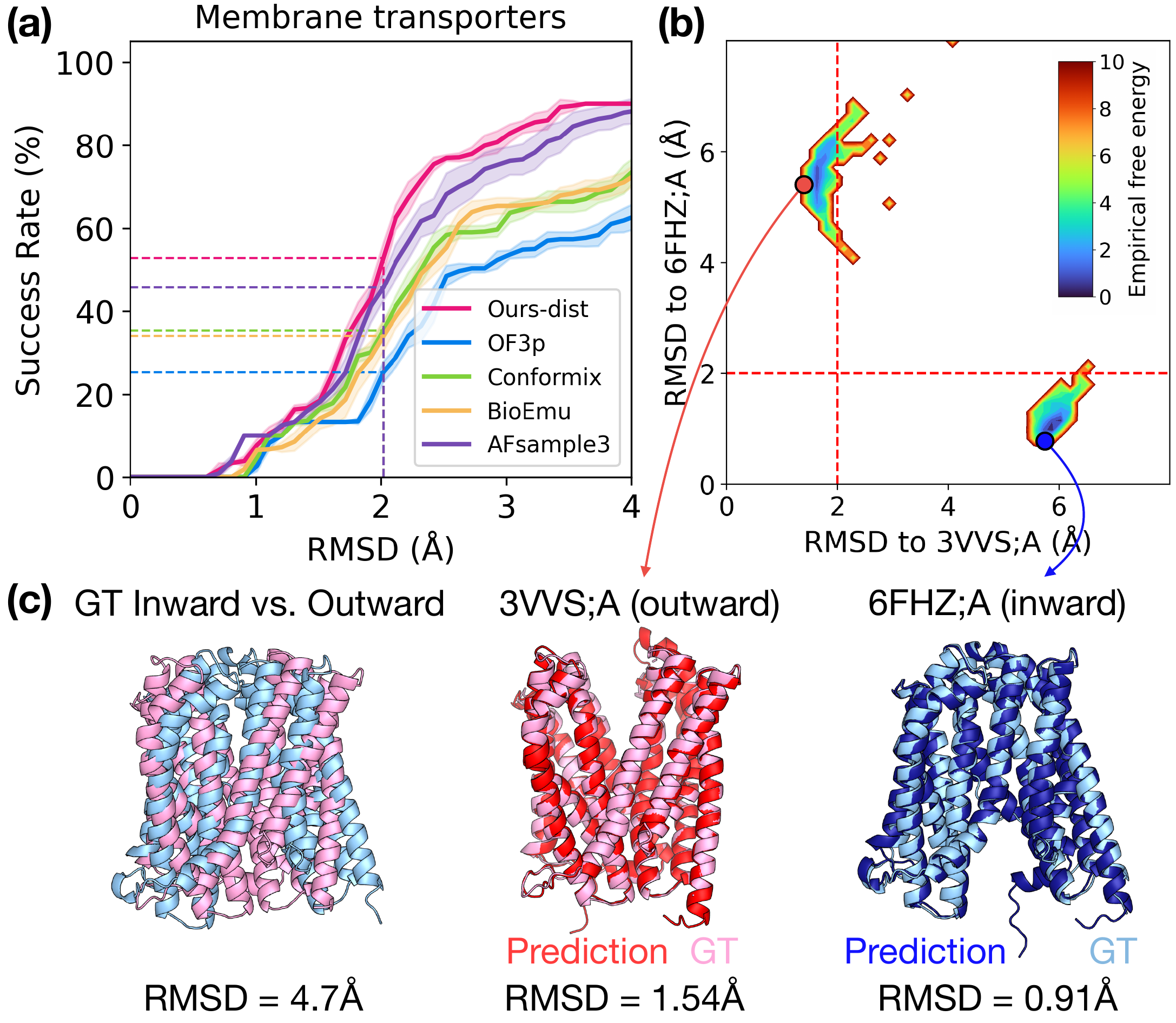}
    \caption{\textbf{State coverage of membrane transporters.}
    \textbf{(a)} Success@100 rate of covering membrane transporter states (see text for definition) as a function of RMSD for ConforNets and other methods. Shaded regions denote bootstrap standard deviation.
    \textbf{(b)} Empirical energy landscape of ConforNet samples as a function of RMSD to inward-open (\href{https://www.rcsb.org/structure/6FHZ}{6FHZ};A) and outward-open (\href{https://www.rcsb.org/structure/3VVS}{3VVS};A) states of MATE family transporter. \textbf{(c)} Best predictions and ground truth conformations superposed.}
    \label{fig:membrane_ex}
\end{figure}

\subsection{Metrics}
\label{sec:metrics}

Multi-state benchmarks \citep{bioemu} evaluate whether a method can generate \emph{any} sample that matches a target state. Following this convention, we define \textbf{success@$\boldsymbol{B}$} as having a sample whose backbone RMSD $\leq\tau$ to the reference state after $B$ predictions. For each state, we estimate success probability using 100 bootstrap trials. The cutoff $\tau$ is determined based on the scale of conformational change observed in a given benchmark (App.~\ref{app:success_cutoff}). Benchmark success rate is the mean success rate across all reference states within the benchmark. We additionally report success rate as a function of $\tau$ to yield coverage curves.

\subsection{Results}
\label{sec:results_main}

ConforNets consistently achieve state-of-the-art success rates across benchmarks in both the coordinate and distogram formulations (Table \ref{tab:main}), with the latter showing the best overall results. One exception is the cryptic pockets benchmark, where the stricter RMSD cutoff of 1Å (on the pocket region) favors the coordinate objective. ConforMix performs competitively on domain motions, fold switchers, and OOD60, but trails on other benchmarks. BioEmu underperforms relative to OF3p-based methods; however, direct comparison is difficult as it was trained on an earlier PDB cutoff (less data) and fine-tuned on MD trajectories (additional synthetic data). Random MSA perturbations, whether along rows (default OF3p or Shallow) or columns (AFsample3), underperform targeted latent space perturbations (ours) that explicitly promote diverse predictions. Still, column masking performs substantially better than row subsampling, presumably because it more directly alters inferred residue-residue contacts.

On membrane transporters, ConforNets show the greatest advantage (Fig.~\ref{fig:membrane_ex}a). While we do not expect ConforNets to be energetically calibrated, we nonetheless computed a ConforNets-based empirical energy landscape\footnote{We compute this landscape by estimating a smoothed 2D density over samples in the space of RMSDs to the two reference conformations, then converting it to a free energy.} for a MATE family transporter. ConforNets more or less exclusively sample the transporter's two reference states with clear funnels (Fig.~\ref{fig:membrane_ex}b) and predict structures for both states that well match their respective ground truths (Fig.~\ref{fig:membrane_ex}c).

\begin{figure}[t!]
    \centering
    \includegraphics[width=\linewidth]{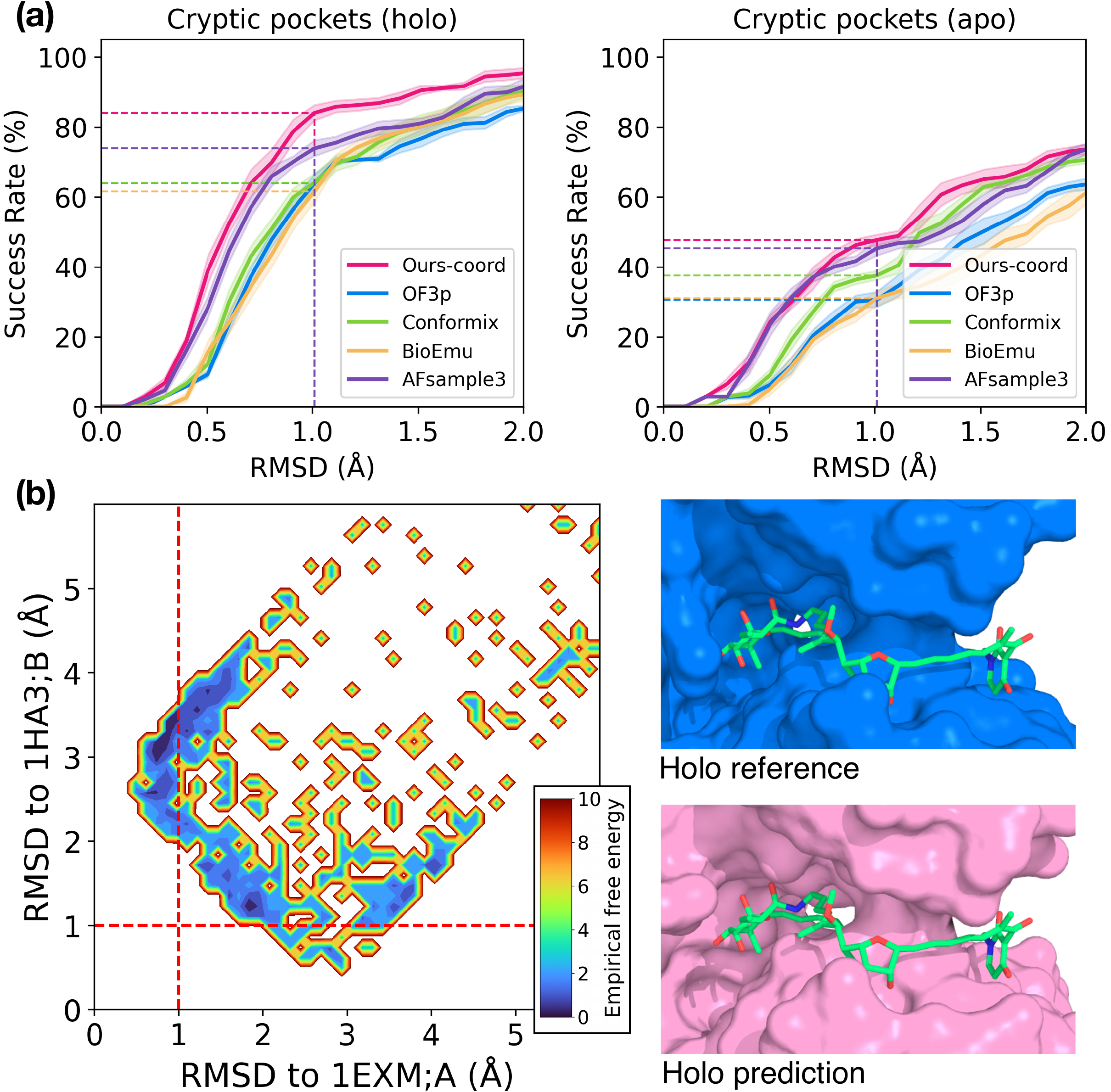}
    \caption{\textbf{State coverage of cryptic pockets.}
    \textbf{(a)} Success@100 rates of covering cryptic pockets in the holo (left) and apo (right) states as a function of RMSD. Shaded regions denote bootstrap standard deviation.
    \textbf{(b)} Empirical energy landscape of ConforNet samples as a function of RMSD to apo (\href{https://www.rcsb.org/structure/1EXM}{1EXM};A) and holo (\href{https://www.rcsb.org/structure/1HA3}{1HA3};B) pockets of Elongation Factor Tu. Right: Experimental (top) and predicted (bottom) structures of the cryptic pocket with N-methyl kirromycin.}
    \label{fig:cryptic_ex}
    \vspace{-10pt}
\end{figure}

Although ConforNets operate globally in latent space, they can induce localized structural changes. On cryptic pockets, coordinate-based ConforNets achieve substantial gains over baselines despite their coordinate MSE loss being defined over the whole structure (Fig.~\ref{fig:cryptic_ex}a). This suggests that latent-space exploration focuses on physically feasible and energetically accessible degrees of freedom. Fig.~\ref{fig:cryptic_ex}b visualizes ConforNets' empirical free energy landscape as a function of pocket RMSD to the apo and holo states for Elongation Factor Tu. ConforNets populate both conformations and produce structures that closely match the holo pocket geometry, as illustrated by the quality of the pocket surface and ligand alignment. Nonetheless, the sampled empirical free energy landscape is clearly not calibrated.

\section{Supervised Transfer of Conformations}
\label{sec:transfer_results}

\subsection{Benchmarks}
\label{sec:transfer_benchmarks}
In this section, we evaluate a ConforNet's ability to transfer a conformation from one protein to another. As this is a novel task, we construct three conformation transfer benchmarks that satisfy two criteria: (i) the conformational change is biologically meaningful and conserved across the protein family, and (ii) the source state being transferred is rarely sampled by the base model (in this case, default OF3p inference); this ensures that success is attributable to controlled induction rather than default sampling.

\begin{itemize}[noitemsep, topsep=0pt]
\item \textbf{GPCR active (N=51):} GPCR activation couples agonist binding to G-protein recruitment by rearranging transmembrane helices (TM) 5--7 \citep{latorraca2017gpcr,hauser2021gpcr}. We curate 51 receptors (App.~\ref{app:gpcr_preprocess}) with experimental inactive and active structures from GPCRdb \citep{kooistra2021gpcrdb} and compute RMSD of TM6 to its undersampled active state.

\item \textbf{Kinase DFG-out (N=20):} Kinases transition between active (DFG-in) and inactive (DFG-out) states by locally rearranging their activation loop and DFG motif. We curate 20 kinase DFG-in/out pairs (App.~\ref{app:kinases_preprocess}) from KLIFS \citep{KLIFS} and compute RMSD of the activation loop and DFG motif to their undersampled inactive state.

\item \textbf{Transporter outward-open (N=15):} We reuse the transporter benchmark (Sec.~\ref{sec:benchmarks}) and compute global RMSD to the undersampled outward-open state.
\end{itemize}

\begin{figure}[t!]
    \centering
    \includegraphics[width=1\linewidth]{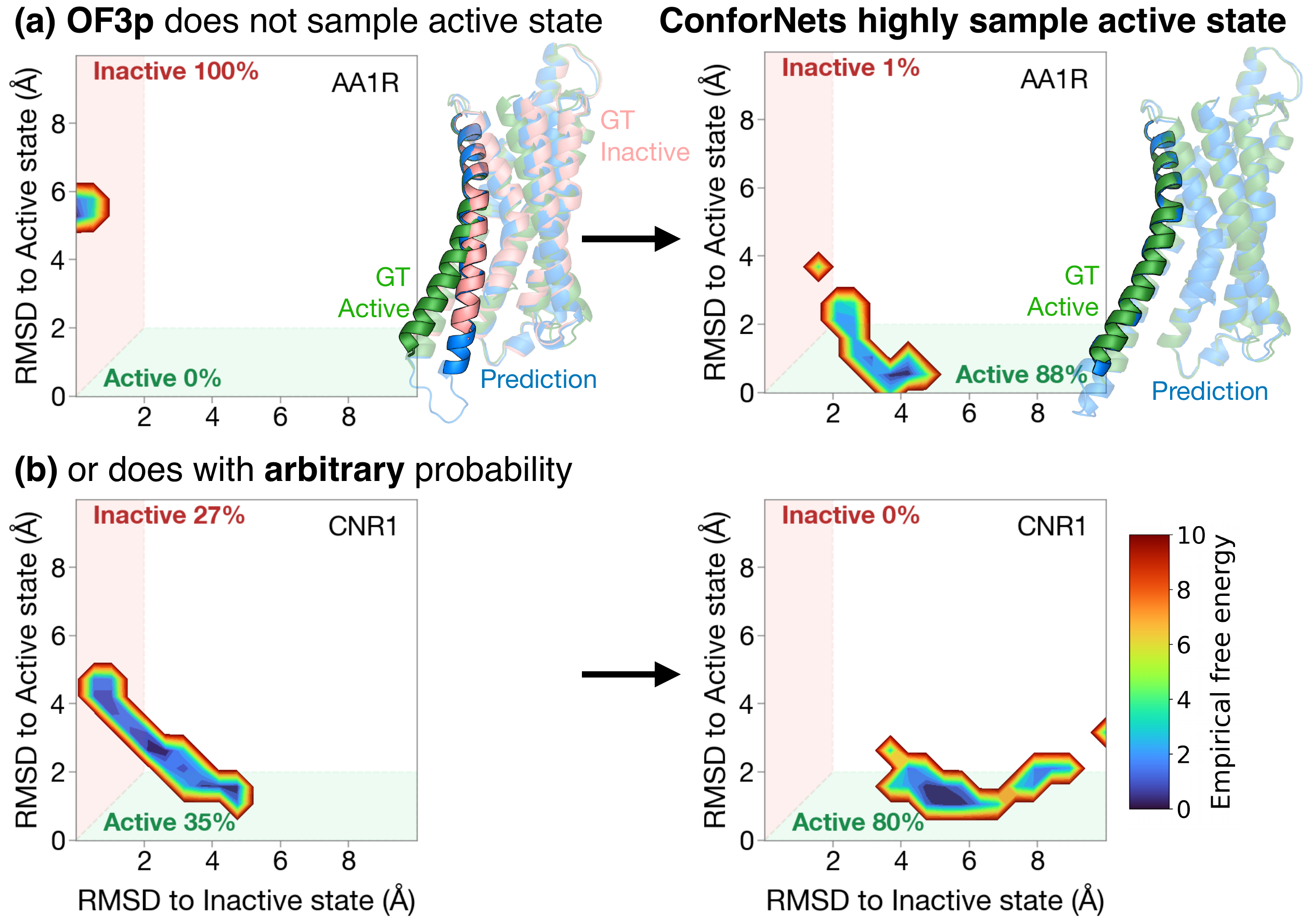}
    \caption{\textbf{Transfer ConforNets shift state distribution.} Empirical energy landscapes of two GPCRs under default OF3p sampling (left) and ConforNets (right). For Adenosine receptor A1 \textbf{(a)}, OF3p never samples the active state, while for Cannabinoid receptor 1 \textbf{(b)}, OF3p samples it with an arbitrary probability of 35\%. ConforNets shift the sampling probabilities of the active states of both GPCRs to over 80\% and rarely sample their inactive states.}
    \label{fig:problem_intro}
\end{figure}

\begin{table*}[th!]
\centering
\caption{\textbf{Conformation transfer evaluated by reachability and at-will induction for three benchmark protein families.} Current methods do not provide an interface for inducing a conformation embodied by a different protein and are therefore shown in \textcolor{gray}{gray} (success@5 should be interpreted as the reference sampling frequency). We only generate 100 samples per benchmark/model combination and therefore do not report standard deviation from bootstrap trials for success@100.} 
\label{tab:transfer_success_5_100}
\small
\begin{tabular}{lcccccc}
\toprule
& \multicolumn{3}{c}{\textbf{At-will Induction} (success@5)} & \multicolumn{3}{c}{\textbf{Reachability} (success@100)} \\
\cmidrule(lr){2-4} \cmidrule(lr){5-7}
Method & GPCR active & Kinase DFG-out & Transp. Out & GPCR active & Kinase DFG-out & Transp. Out \\
\midrule
OF3p
& \textcolor{gray}{\bmcell{24.3}{2.6}} & \textcolor{gray}{\bmcell{5.9}{2.3}} & \textcolor{gray}{\bmcell{16.1}{5.8}}
& 37.3\% & 10.0\% & 33.3\% \\

ConforMix
& \textcolor{gray}{\bmcell{16.9}{3.7}} & \textcolor{gray}{\bmcell{3.7}{3.6}} & \textcolor{gray}{\bmcell{23.1}{7.0}}
& 43.1\% & 15.1\% & 40.0\% \\

AFsample3
& \textcolor{gray}{\bmcell{27.4}{4.1}} & \textcolor{gray}{\bmcell{4.4}{3.7}} & \textcolor{gray}{\bmcell{20.2}{5.7}} 
& 60.8\% & 20.0\% & 33.3\% \\

Template
& \bmcell{14.9}{3.0} & \bmcell{6.4}{2.6} & \bmcell{15.7}{6.6}
& 32\% & 10.5\% & 35.7\% \\

ConforNets
& \bmcell{\textbf{79.1}}{2.3} & \bmcell{\textbf{22.8}}{3.8} & \bmcell{\textbf{56.7}}{6.5}
& \textbf{86.0}\% & \textbf{26.3}\% & \textbf{73.3}\% \\
\bottomrule
\end{tabular}
\end{table*}

\subsection{Metrics}
\label{sec:transfer_metrics}
We consider two metrics: \textbf{success@100} as a test of reachability, \textit{i.e.}, whether the desired state can be generated at all, with low success rates indicating that it is rarely or never sampled (Fig.~\ref{fig:problem_intro}a); and \textbf{success@5} as the indicator of at-will induction, since the goal of transfer is to reliably induce a specific functional state (Fig.~\ref{fig:problem_intro}b). To declare a sample successful we use an RMSD cutoff of $\tau=2$\AA{}. For GPCRs and kinases, we first align the full structures (using US-align \citep{zhang2022us}) and then compute RMSD over the regions involved in the conformational change (Sec.~\ref{sec:transfer_benchmarks}). We estimate the probability of success@5 via 100 bootstrap trials. The benchmark transfer success rate is the average across all target proteins in the benchmark (excludes the source protein used in training the ConforNet).

\subsection{Experimental design}
For each family, we choose as the source protein the family centroid, defined as the protein with the highest mean sequence similarity to all other family members. We assess the suitability of our heuristic for source selection in App.~\ref{app:source_selection}. We train 10 ConforNets on this source and apply them to every other protein in the benchmark, generating 10 diffusion samples per ConforNet, yielding 100 samples per target. ConforNets were trained for at most 300 steps with early stopping if the loss fell below 0.1 for three consecutive steps; the checkpoint with the minimum loss was used. As baselines, we ran default OF3p, OF3p with the centroid protein's desired conformation provided as a template, and AFsample3, each with 20 random seeds, as well as ConforMix with 1 seed. All four settings yield 100 samples per target. In all settings, we used $R=1$ pass which performed similarly or better than $R=11$ for baselines (App.~\ref{app:recycle_transfer_baseline}).

\begin{figure}[t!]
    \centering
    \includegraphics[width=1\linewidth]{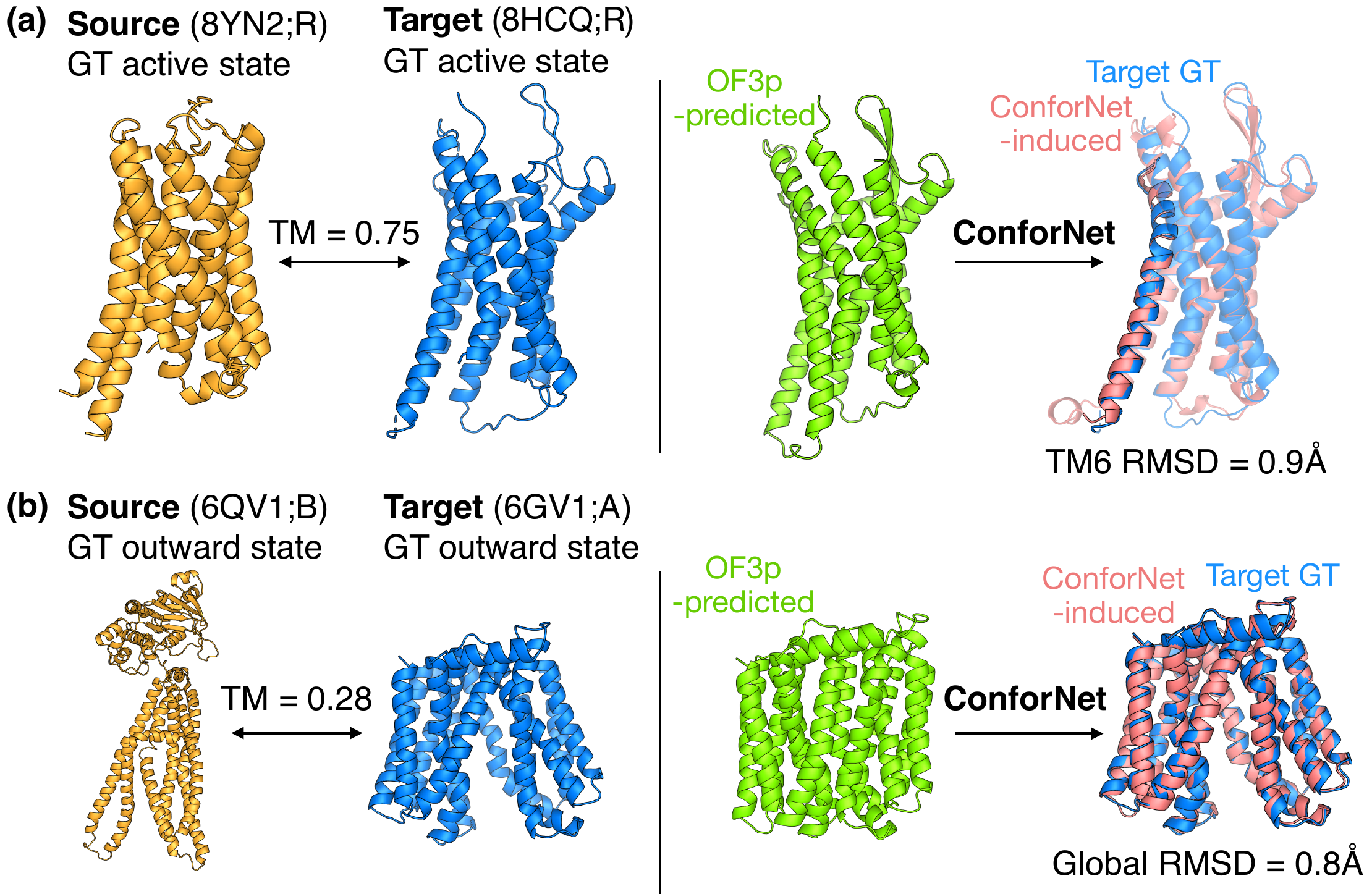}
    \caption{\textbf{Conformation transfer generalizes to structurally dissimilar proteins}. ConforNets trained on source proteins with desired conformations (yellow) can be applied to target proteins whose default OF3p prediction is of a different state (green) and whose global structural similarity with respect to the source---in the desired state---is low (blue). Nonetheless, ConforNet-induced predictions (pink) are highly congruent with the ground truth target structures (blue). \textbf{(a)} GPCR active state transfer with RMSD reported over highlighted TM6 region. \textbf{(b)} Membrane transporter outward-open-state transfer with global backbone RMSD shown.}
    \label{fig:transfer_dissimilar}
\end{figure}

\subsection{Results}

Across our three benchmarks, ConforNets trained on a single source protein substantially increase the sampling frequency of its respective conformation over default OF3p inference (Table~\ref{tab:transfer_success_5_100}). Supplying the desired conformation of the centroid protein as a template does not improve performance. Low reachability, even when using competitive diversity-maximizing baselines such as ConforMix and AFsample3, highlights that supervised transfer enables a new capability. In absolute terms, kinases are the most challenging, possibly due to the difficulty of modeling flexible loops compared to the well-ordered secondary structures of GPCRs and transporters.

Transfer success only weakly depends on source-target similarity (Fig.~\ref{fig:transfer_similarity}b). Fig.~\ref{fig:transfer_dissimilar} illustrates successful transfer for a target with a TM-score of 0.28 to the ConforNet source protein. For reference, OF3p does not produce a single sample that is close to the active state shown in Fig.~\ref{fig:transfer_dissimilar}a (minimum TM6 RMSD of 3.34\AA{}), while for the outward state shown in Fig.~\ref{fig:transfer_dissimilar}b, ConforNets quintuple the fraction of successful predictions, from 7\% to 35\%.

\begin{figure}[t!]
    \centering
    \includegraphics[width=1\linewidth]{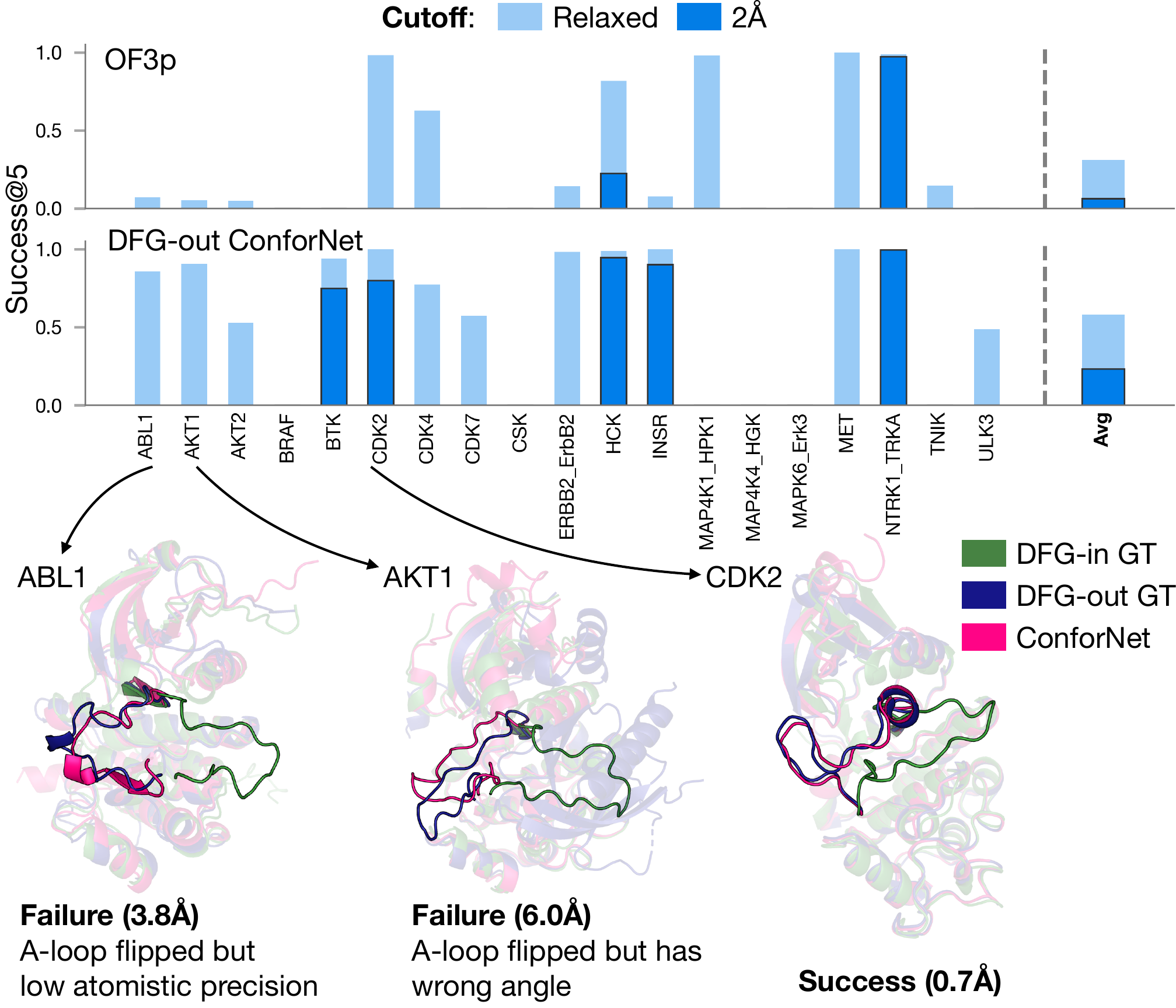}
    \caption{\textbf{Success and failure modes of kinase transfer.} Bar plots showing average (excluding source protein) and per-target success@5 rates for OF3p and ConforNet-induced transfer of the kinase DFG-out state using 2Å RMSD (blue) and relaxed (light blue) cutoffs. Three illustrative examples of transfer outcomes are shown as structural overlays of the ConforNet prediction (pink) with ground truth DFG-in (green) and DFG-out (purple) states.}
    \label{fig:transfer_kinases}
\end{figure}

We next analyze the success and failure modes of ConforNets vs. those of OF3p (Fig.~\ref{fig:transfer_kinases}). Kinase activation loops (A-loop) can undergo $\sim$180° flips, leading to large deviations in RMSD (up to 20Å) over the DFG motif and A-loop region. Given the scale of change, we consider a relaxed criterion for detecting the A-loop flip\footnote{Of the 20 kinases in the benchmark, only 11 have experimentally resolved A-loop flips between the pairs (App.~\ref{app:kinases_preprocess}).} (light blue vs. blue in Fig.~\ref{fig:transfer_kinases}). It asserts that a prediction is closer to the DFG-out than the DFG-in state, and its RMSD to the DFG-out state is less than $\max[2, \tfrac{1}{2}\mathrm{RMSD}(\text{DFG-in}, \text{DFG-out)}]$. Under both criteria we see improvements over OF3p sampling (6\%$\to$23\% and 30\%$\to$58\% for strict and relaxed, respectively). Comparing the two criteria reveals distinct failure patterns: in many cases, OF3p fails to generate conformations that approach the inactive state, with 0\% success rate even with the relaxed criterion. In contrast, ConforNets often produce structures closer to the inactive state, but with limited precision or angular deviations.

\section{Discussion}

\subsection{Optimal perturbation location}
\label{sec:results_perturbation}

\begin{table}[t]
\centering
\caption{Mean and standard deviation of RMSD between the ConforNet-induced predictions and the ground truth entries of OOD60 using different perturbation locations. Statistics are computed across 38 entries $\times$ 4 replicates. ConforNets are trained directly to ground truths using $K=1$ mini rollouts.}
\label{tab:gt_sweep_main}
\begin{tabular}{lcc}
\toprule
Latent & Mini RMSD & Full RMSD \\
\midrule
$\mathbf{z}^{\mathrm{post}}$ & 1.79 $\pm$ 1.19 & 3.40 $\pm$ 2.37 \\
$\mathbf{z}^{\mathrm{pre}}$  & 1.90 $\pm$ 2.05 & \textbf{1.93 $\pm$ 1.55} \\
$\mathbf{s}^{\mathrm{post}}$ & 2.31 $\pm$ 2.99 & 9.84 $\pm$ 26.33 \\
$\mathbf{s}^{\mathrm{pre}}$  & 4.14 $\pm$ 3.78 & 4.41 $\pm$ 3.85 \\
\bottomrule
\end{tabular}
\end{table}

Our choice of applying ConforNets to the pre-Pairformer pair representation ($\mathbf{z}^{\mathrm{pre}}$) is intentional---we arrived at this choice through experiments that varied ConforNet application along two axes: (i) single vs. pair representations and (ii) pre- or post-Pairformer. In aggregate, we evaluated ConforNets on $\mathbf{z}^{\mathrm{pre}}$, $\mathbf{z}^{\mathrm{post}}$, $\mathbf{s}^{\mathrm{pre}}$, and $\mathbf{s}^{\mathrm{post}}$. For each latent, we optimized 4 ConforNets for the ground truth entries of OOD60 ($N$=19$\times$2) using $\mathcal{L}_{\text{transfer}}$ and $R=1$. We used this setup not to transfer states, but to determine whether a given latent provides sufficient conformational control to induce a state that is directly optimized for. While perfect recovery cannot be expected due to stochasticity in MSA and diffusion sampling, the ideal perturbation location should achieve low RMSD across all targets after optimization.

Finally, we varied the number of diffusion steps $K \in \{1,2,5,10\}$ used to predict structures for computing the loss, which we refer to as a \emph{mini rollout}. After optimization, we evaluated RMSD to the ground truth target under two settings: (i) using the same $K$-step mini rollout and (ii) using a full rollout of 200 steps. Table~\ref{tab:gt_sweep_main} summarizes the $K=1$ results (full sweep is in App. Table~\ref{tab:gt_sweep_full}). We find that $\mathbf{z}^{\mathrm{pre}}$ achieves consistently low RMSD under both mini and full rollouts. In contrast, post-Pairformer ConforNets, especially $\mathbf{s}^{\mathrm{post}}$, fit the mini-rollout but degrade under full rollout, indicating potentially unstable perturbations that shortcut the diffusion process. ConforNets on $\mathbf{s}^{\mathrm{pre}}$ fail to achieve low RMSD, indicating they lack sufficient conformational control.

\subsection{Compute overhead}
\label{sec:discussion_time}

Predictions made using trained ConforNets incur negligible additional cost. Training ConforNets does have a one-time cost, but it can be amortized across many diffusion samples (in both the transfer and diversity-maximizing settings). In terms of wall-clock time, generating 5 diffusion samples in the diversity-maximizing setting costs $2$--$3\times$ default OF3p sampling, similar to ConforMix (Table~\ref{tab:wallclock_time}). In general, the overhead will depend on several implementation choices (App.~\ref{app:compute_overhead}). This overhead is justified as ConforNets change the underlying conformational distribution while simply generating 2$\times$ more OF3p samples does not.

\subsection{Analysis of learned ConforNets}
\label{sec:analysis_weights}

We analyze ConforNets' learned affine transform parameters $\mathbf{W}$ and $\mathbf{b}$ in the context of the transporters benchmark, where we evaluate both diversity and transfer. We observe channel mixing, measured by the norm of the off-diagonal entries of $\mathbf{W}$, and channel scaling, measured by the deviation of the diagonal entries from identity (Fig.~\ref{fig:wb_weights}). Although the off-diagonal entries of $\mathbf{W}$ are small in absolute magnitude, performance drops when we prevent channel mixing, either by removing $\mathbf{W}$ or by restricting it to be a diagonal matrix (Table~\ref{tab:membrane_affine_ablation}). This suggests that even weak cross-channel mixing can propagate through the Pairformer to meaningfully alter the final pair representation (App.~\ref{app:foldswitch_interp}). We also find that ConforNets trained for transfer exhibit larger magnitudes and broader distributions in both $\mathbf{W}$ and $\mathbf{b}$, potentially because they are trained for longer. We discuss these results further in App.~\ref{app:weights_extended}.

\section{Conclusion}

In this work, we introduce ConforNets, a lightweight perturbational approach that enables a new form of conformation control in AF3-based models. Clear opportunities lie in exploring its use in biological applications. For instance, predicting GPCR active conformations can facilitate docking of ligands, while sampling more conformationally diverse ensembles, even if uncalibrated, can provide better starting points for MD simulations. Our study also raises questions regarding the mechanism and limits of conformational transfer. We evaluated transfer primarily at the level of global fold and medium- to large-scale conformational changes, rather than detailed residue-level interactions. Understanding the drivers of transferability, and whether it can be extended to finer structural features, is an important future direction. More broadly, ConforNets could be extended to satisfy diverse constraints---such as ones coming from experimental sources---or to induce distinct binding pockets for challenging problems, such as protein-ligand complex prediction.



\clearpage
\section*{Acknowledgements}

We thank Minhuan Li, Luhuan Wu, Julia Rogers, and Jaeyeon Kim for insightful discussions and feedback. We acknowledge funding from the National Institutes of Health (R35-GM150546 and R01-LM014674 to MA). This work used resources of the National Energy Research Scientific Computing Center (NERSC), a Department of Energy User Facility using NERSC award DDR-ERCAP0037622. 

\section*{Impact Statement}

This paper presents work whose goal is to advance the field of protein biology and structure prediction. There are many potential societal consequences of our work, none that we feel must be specifically highlighted here.

\bibliography{main}
\bibliographystyle{icml2026}

\newpage
\appendix
\onecolumn

\renewcommand{\thefigure}{A\arabic{figure}}
\setcounter{figure}{0}
\renewcommand{\thetable}{A\arabic{table}}
\setcounter{table}{0}

\section{Baselines}
\label{app:baselines}

In this section, we provide additional details on the baselines.

\subsection{ConforMix}

We followed the paper and the official codebase. We used the default parameters of the codebase: a twist strength of 15 and structured-regions-only guidance. We swept the target RMSD from 0 to 20 as in the paper; more precisely, we swept from 0.5 to 19.5 with a step size of 1.0, totaling 20 steps. 

While the official codebase uses a single seed to generate the Pairformer representations and the baseline prediction, we used 8 seeds, since MSA subsampling introduces stochasticity in both representations and structures. To be fair with the other methods that may benefit from this stochasticity for diversity, we ran ConforMix with 8 seeds, 20 target RMSDs, and 5 diffusion samples, resulting in 800 structures per test case.

\subsection{BioEmu}
Since RMSD is computed on the backbone, we did not reconstruct side chains for BioEmu generations, nor did we perform structure relaxation. We generated 4,000 structures per test case without any generation-time filtering.

\subsection{AFsample3}
We adapted the official \href{https://github.com/wallnerlab/afsample3}{AFsample3 implementation} to OF3p, in which input MSA columns are randomly replaced with X (unknown) residues under a uniform mutation probability. We used the two best-performing masking levels reported in the paper, 0.2 and 0.4.

\section{Inputs for structure prediction}
\label{app:input_query}

Here, we detail the inputs for structure prediction.

\begin{itemize}
    \item \textbf{MSA}: We query the ColabFold MSA server \citep{mirdita2022colabfold} with the standard setting in OF3p.
    \item We did not provide the \textbf{templates}, as they would bias structure prediction.
    \item \textbf{Sequence}: Following BioEmu, when the sequences of two reference structures differ, we sample both sequences in equal proportion. However, for OOD60 we observed that the mismatch was primarily due to His-tags and therefore sampled only the de-tagged sequence. 
    \item \textbf{Ligand inputs} for the cryptic pocket benchmark: Although OF3p supports ligand conditioning, we evaluate all methods without ligands for direct comparison with BioEmu, which does not support ligand inputs, and because the apo state is the less represented conformation. Most methods achieve lower success on apo than on holo conformations, likely due to a PDB deposition bias toward holo complexes. We analyze the effect of providing ligands in App.~\ref{app:holo_query}.
\end{itemize}

\clearpage 

\section{Multi-state benchmarks}

In this section, we provide additional details on the multi-state benchmarks.

\subsection{RMSD cutoff $\tau$}
\label{app:success_cutoff}
When computing success, we use the benchmark-specific RMSD cutoff $\tau$ to account for the scale of the conformational change. 
For OOD60 and domain motions, which involve large domain-level rearrangements, we adopt $\tau = 3$\AA{} following BioEmu. For cryptic pockets, where conformational changes are localized and apo/holo pairs differ by as little as 1.02\AA{}, we use a stricter cutoff of $\tau = 1$\AA{}. For both OOD60 and the cryptic pockets, RMSD is computed over the curated local regions provided by BioEmu (the pocket region for the cryptic pockets), which we follow exactly.

ConforMix instead defines success relative to the inter-conformation distance: a prediction is considered successful if its RMSD to one reference state is less than half the RMSD between the two reference conformations. While adaptive, this criterion can be overly permissive for benchmarks with large conformational changes (\textit{e.g.}, when the two states differ by $>$20\AA{}). We therefore use fixed RMSD thresholds for the remaining benchmarks: $\tau = 3$\AA{} for fold switchers and $\tau = 2$\AA{} for membrane transporters, where the reference states typically differ by only 4-5\AA{}. See Table~\ref{tab:rmsd_cutoffs} for the summary.

\begin{table}[h]
\centering
\caption{RMSD cutoffs used for success evaluation.}
\label{tab:rmsd_cutoffs}
\small
\begin{tabular}{lcc}
\toprule
Benchmark & RMSD cutoff $\tau$ (\AA) & RMSD region \\
\midrule
Cryptic pockets & 1.0 & Pocket region (BioEmu) \\
Domain motions & 3.0 & Full structure \\
OOD60 & 3.0 & Curated local regions (BioEmu) \\
Membrane transporters & 2.0 & Full structure \\
Fold switchers & 3.0 & Full structure \\
\bottomrule
\end{tabular}
\end{table}

\subsection{Filtering of trivially unphysical samples} 
To prevent under-reporting success due to invalid structures that can be trivially filtered out, we discard any samples failing the BioEmu geometric validity checks for chain continuity and steric clashes (C$\alpha$-C$\alpha < 4.5$\AA, C-N $< 2.0$\AA, clash distance $> 1.0$\AA). Applying our diversity objective did not increase the proportion of unphysical samples compared to OF3p.

\subsection{Software}
We use the \href{https://github.com/microsoft/bioemu-benchmarks}{\texttt{bioemu\_benchmarks}} package released with BioEmu \citep{bioemu} as our evaluation pipeline, with some minor modifications to input formatting, RMSD cutoffs, and success rate computation for different numbers of samples. RMSD is computed following their pipeline: generated structures and reference structures are loaded as trajectories, sequences are aligned, backbone atoms are selected, and RMSD is computed using \texttt{mdtraj} \citep{McGibbon2015MDTraj}.

\clearpage

\section{Conformation transfer benchmarks}
\label{app:new_benchmarks}

In this section, we describe the preprocessing of conformation transfer benchmark introduced in the paper.

\subsection{Preprocessing active-inactive pairs from GPCRdb}
\label{app:gpcr_preprocess}
The GPCR benchmark was constructed from the GPCRdb structure database \cite{munk2016gpcrdb}. We selected GPCRs with both fully active (100\% activation degree) and inactive structures available. We selected pairs with the fewest engineered mutations and highest crystallographic resolution, retaining 51 pairs where both structures contained at most one mutation. Per-residue segment annotations (TM1-H8) were retrieved from the GPCRdb API, and the canonical 7TM subsequence---spanning TM1 through TM7 including intracellular and extracellular loops---was used as the query sequence for each receptor. Active and inactive structures were downloaded from PDB, and each chain was trimmed to the 7TM domain by global pairwise sequence alignment to the canonical sequence. Inactive structures were further post-processed to remove ICL3 and any fusion proteins inserted, as well as helix 8. Disconnected fragments shorter than 20 residues were discarded. Conformational prediction accuracy was evaluated on the TM6 helix. We do not score other canonical conformational changes, such as TM5 and TM7.

\subsection{Preprocessing DFG in-out pairs from KLIFS}
\label{app:kinases_preprocess}
The kinase benchmark dataset was constructed from the KLIFS database \cite{KLIFS}. We selected kinase families with matching active or inactive state annotations at both the DFG site and the AC-helix. For each kinase family, we selected the pair with the highest sequence similarity, yielding 42 candidate kinase pairs. We filtered out pairs that were missing the canonical DFG sequence motif or that had $<$2\AA{} DFG-loop RMSD between pairs, and we allowed no mismatches between the query and reference sequences over the scored region, yielding 20 pairs. While all pairs captured movement in the DFG site, only 11 pairs had experimentally resolved A-loops and modeled A-loop flipping. These include ABL1, AKT1, AKT2, CDK2, CDK4, CDK7, ERBB2\_ErbB2, MAP4K1\_HPK1, MET, TNIK, and ULK3. All metrics for kinases were scored over the DFG site and A-loop. We did not score other canonical conformational changes, such as AC-helices. All entries were downloaded from the PDB. 

\clearpage

\section{Additional results on multi-state benchmarks}
\label{app:extended_results}

\subsection{Sanity check against the AF3 server}

\begin{figure}[h]
    \centering
    \begin{subfigure}[t]{0.32\linewidth}
        \centering
        \includegraphics[width=\linewidth]{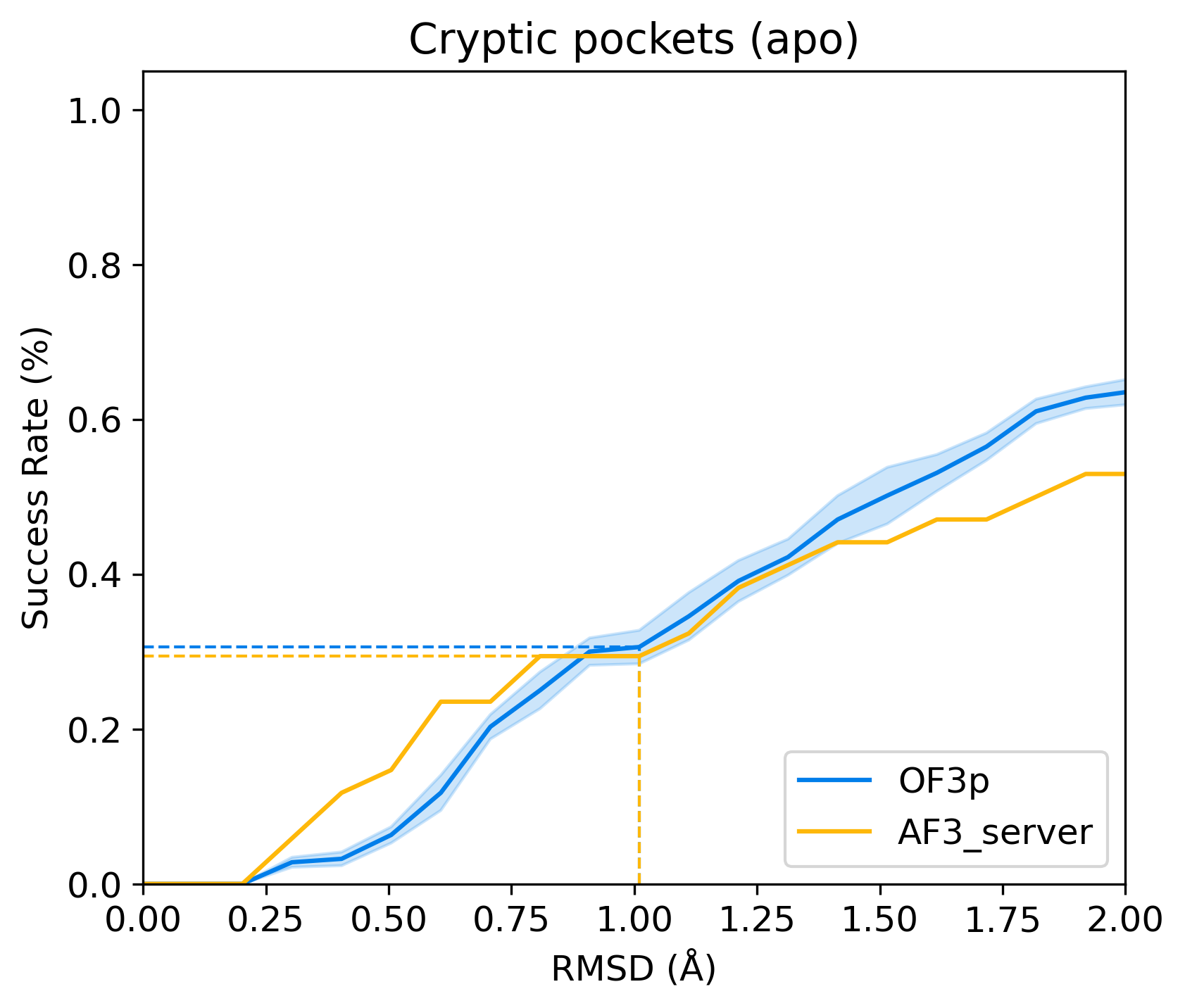}
        \caption{Cryptic pocket (apo)}
    \end{subfigure}\hfill
    \begin{subfigure}[t]{0.32\linewidth}
        \centering
        \includegraphics[width=\linewidth]{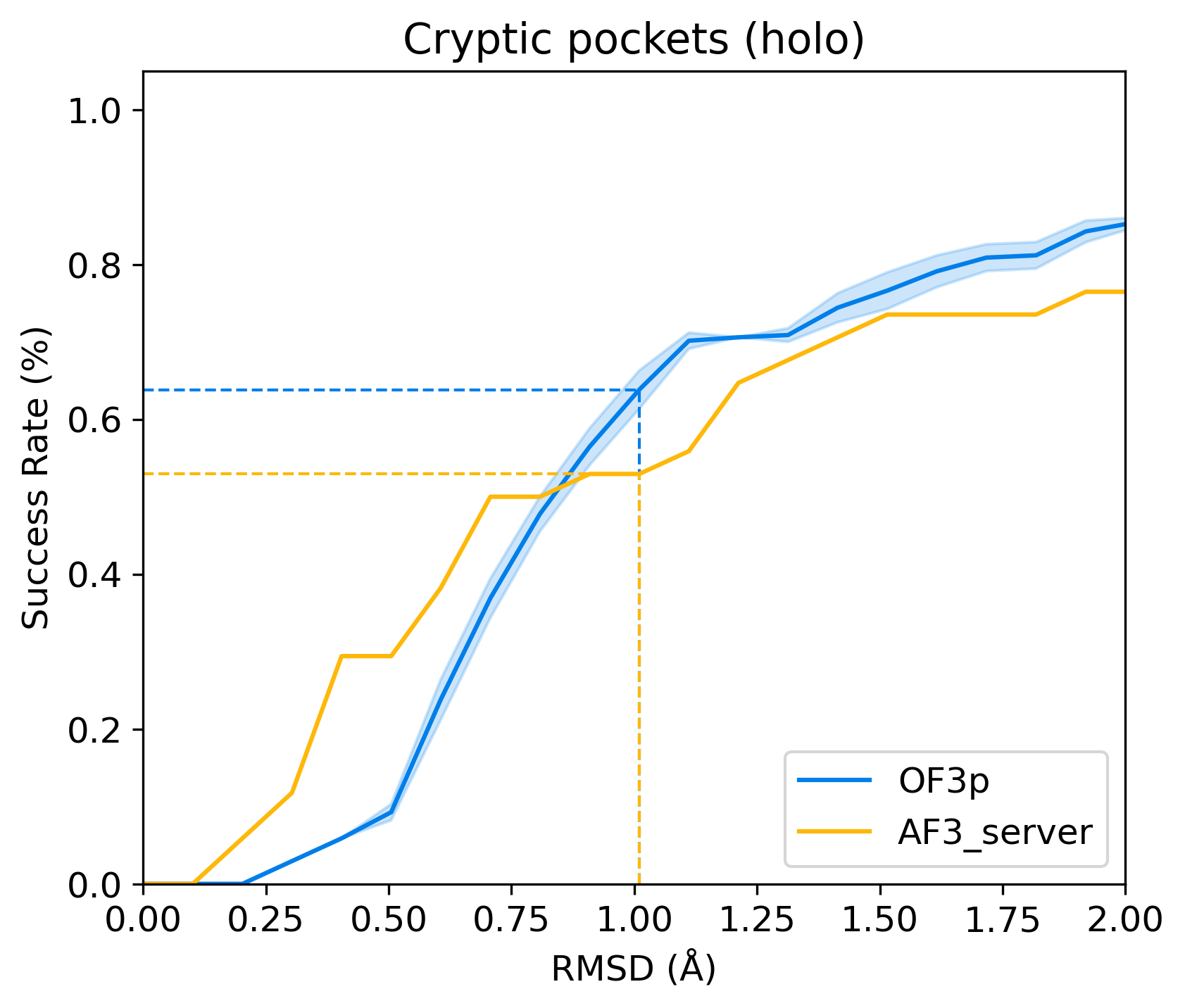}
        \caption{Cryptic pocket (holo)}
    \end{subfigure}\hfill
    \begin{subfigure}[t]{0.32\linewidth}
        \centering
        \includegraphics[width=\linewidth]{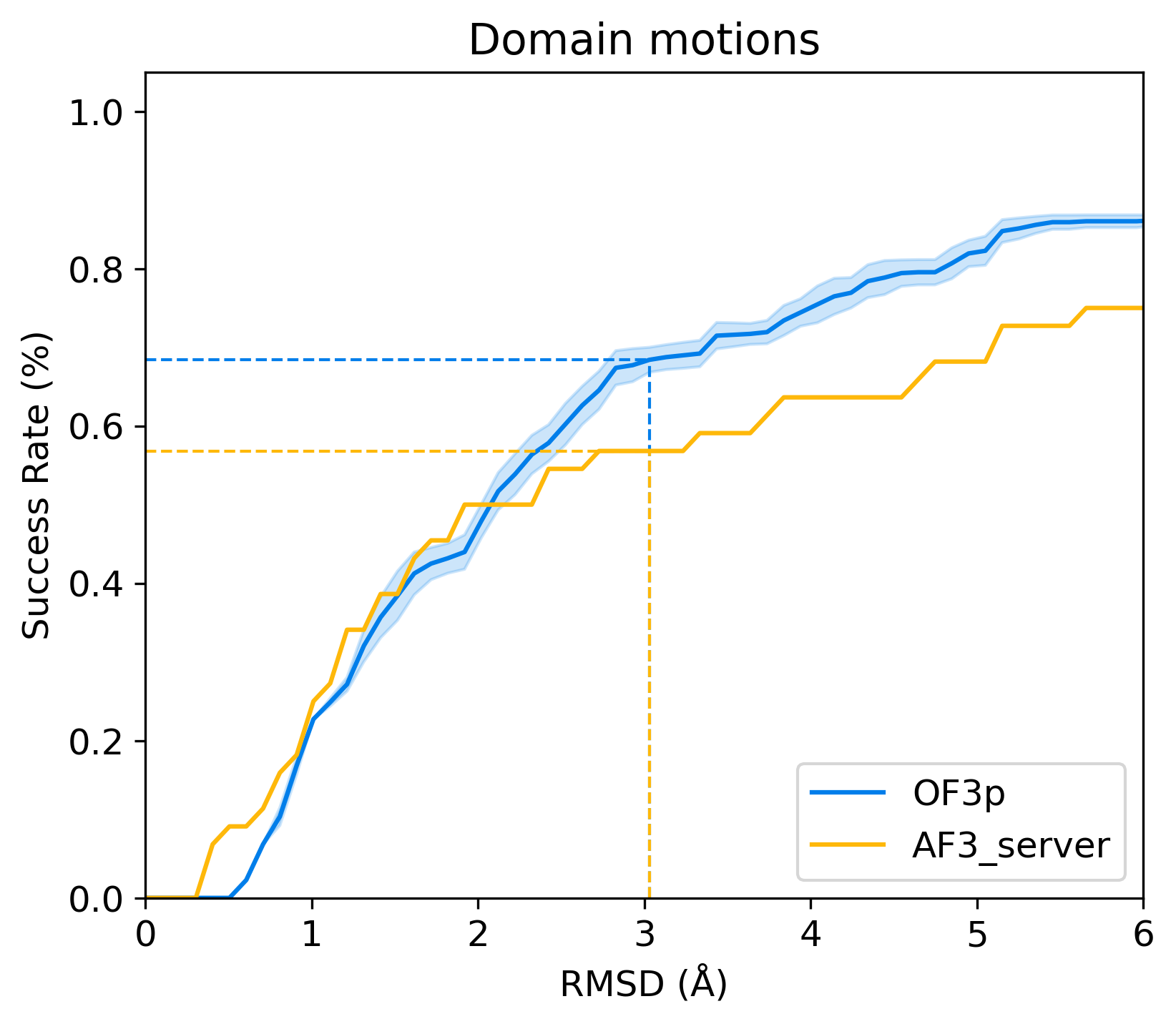}
        \caption{Domain motion}
    \end{subfigure}

    \vspace{0.5em}

    \begin{subfigure}[t]{0.32\linewidth}
        \centering
        \includegraphics[width=\linewidth]{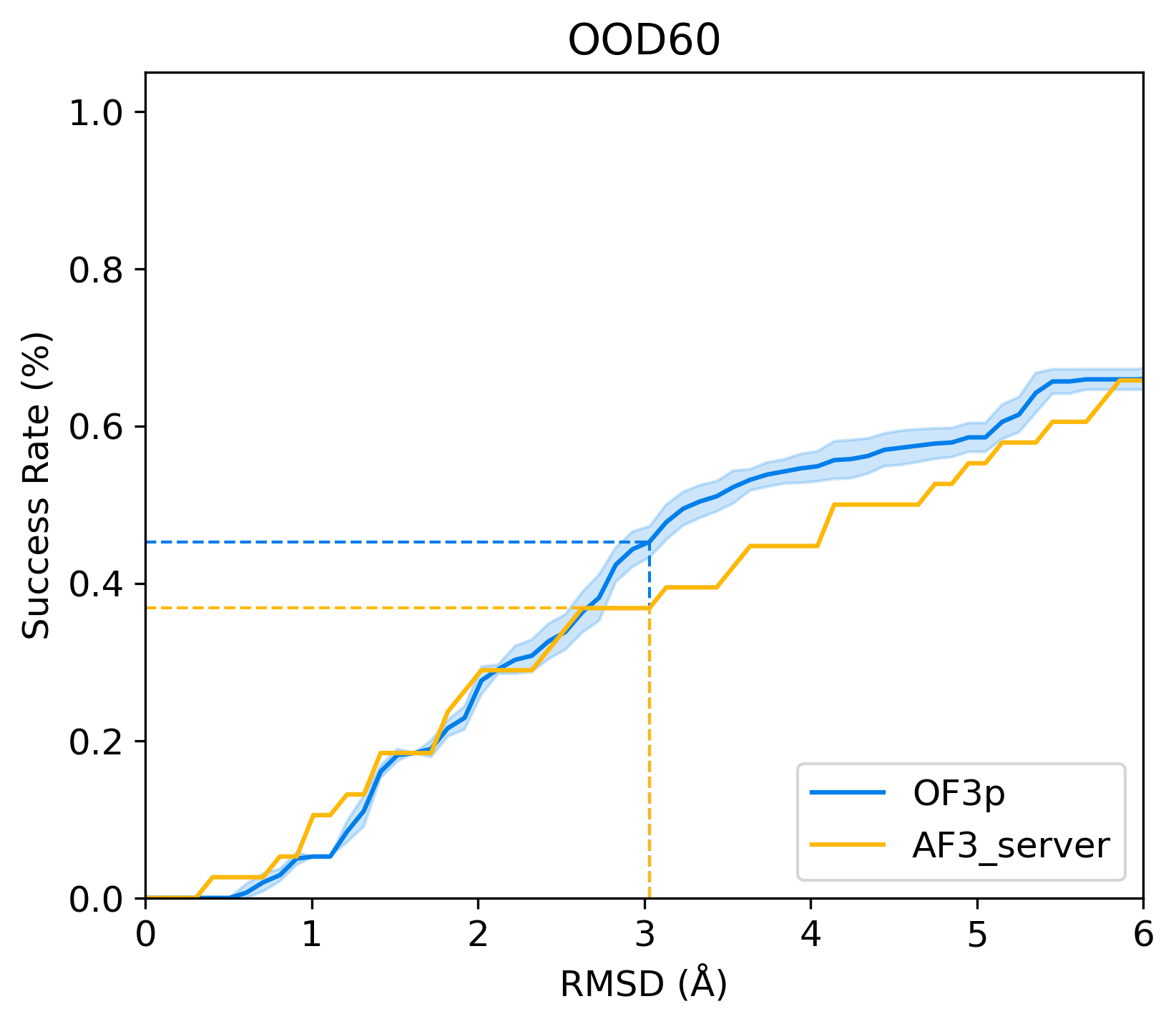}
        \caption{OOD60}
    \end{subfigure}\hfill
    \begin{subfigure}[t]{0.32\linewidth}
        \centering
        \includegraphics[width=\linewidth]{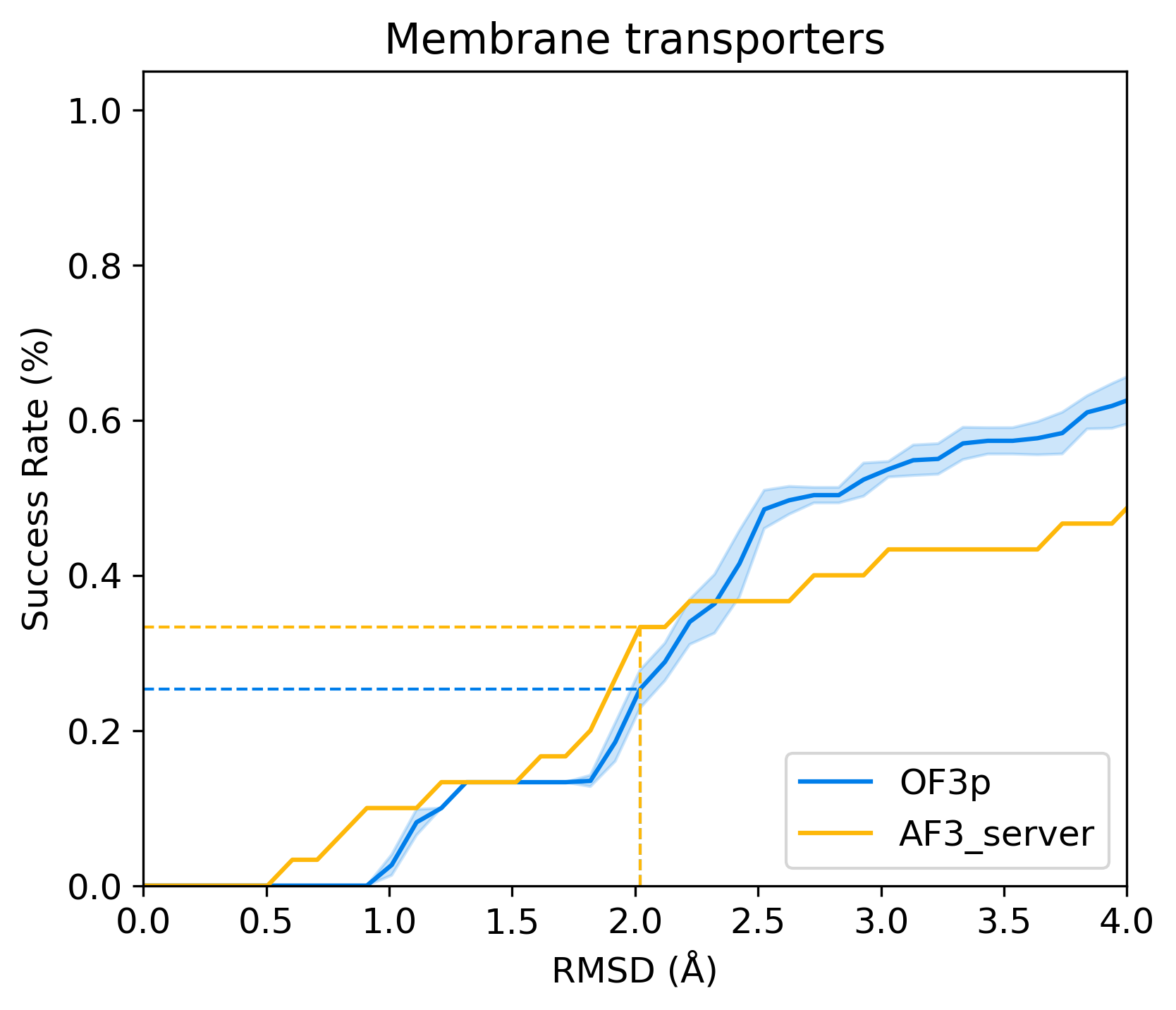}
        \caption{Membrane transporter}
    \end{subfigure}\hfill
    \begin{subfigure}[t]{0.32\linewidth}
        \centering
        \includegraphics[width=\linewidth]{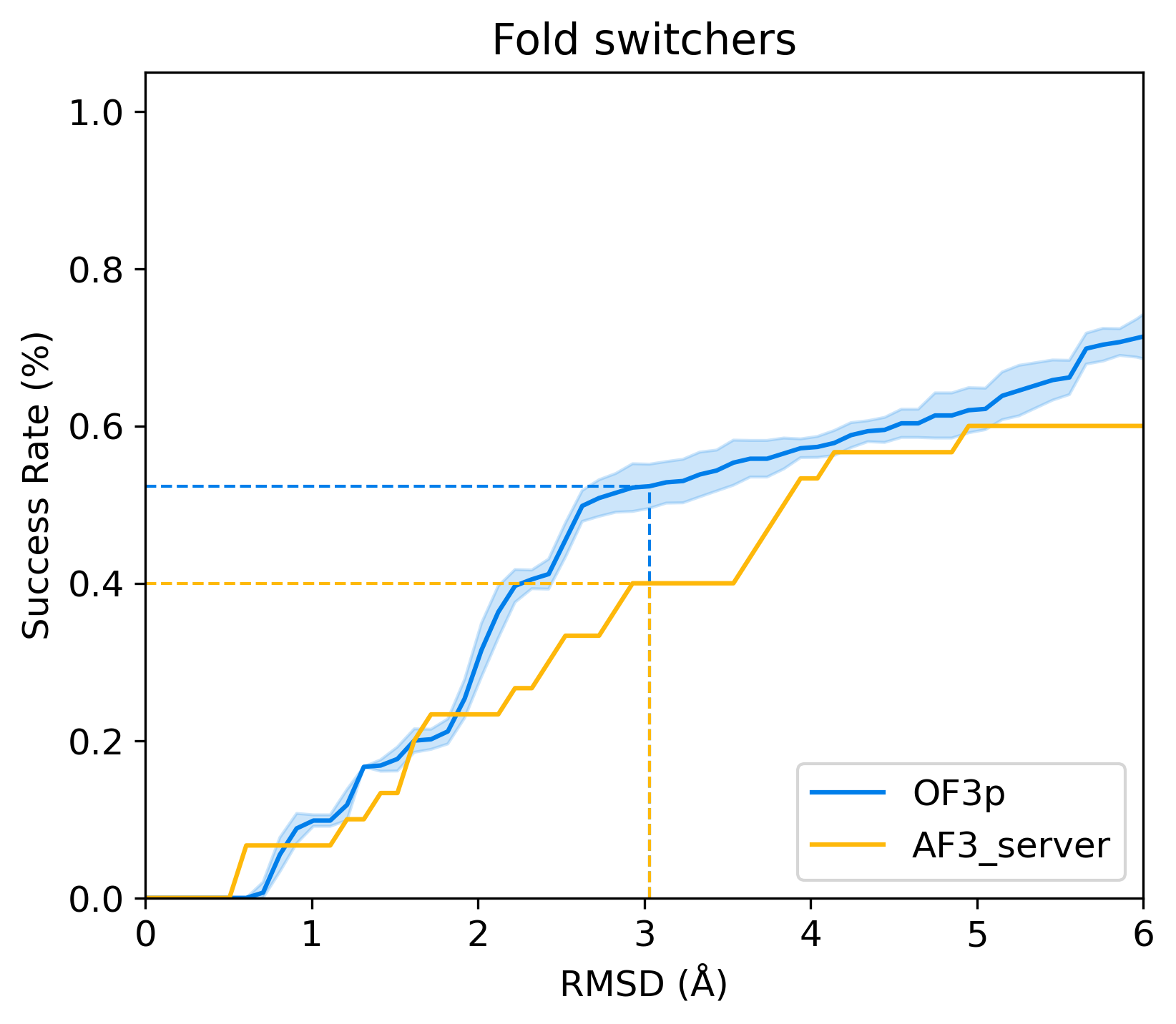}
        \caption{Fold switching}
    \end{subfigure}

    \caption{
        Success@5 rates of covering multi-state benchmarks as a function of RMSD. Shaded regions denote bootstrapped standard deviation for OF3p; AF3 curves are from a single server run.}
    \label{fig:multiconf_af3}
\end{figure}

Since our experiments are conducted on OF3p, an open-source reproduction of AF3, one may question how closely it matches the official AF3 performance. To assess this, we ran one AF3 seed (5 diffusion rollouts) using the AF3 server and report success@5 for both models. Because of the limited API quota, we cannot compute bootstrapped standard deviations for AF3. Accordingly, Table~\ref{tab:comp_af3} and Fig.~\ref{fig:multiconf_af3} should be interpreted only as a sanity check that OF3p performs similarly to AF3, rather than as evidence that OF3p outperforms AF3.

\begin{table*}[htbp]
\centering
\caption{\textbf{Multi-state benchmarks performance of the AF3 server and OF3p.} For OF3p, we report means and standard deviations of success@5 (using 100 bootstrap trials). AF3 results are based on a single server run (one random seed, 5 diffusion rollouts) due to limited API quota, so no bootstrap estimates or standard deviations are available.}
\small
\label{tab:comp_af3}
\begin{tabular}{lccc}
\toprule
Benchmark & $N$ & AF3 server & OF3p \\
\midrule
Cryptic pockets (apo) & 34 & 29.4 & \bmcell{30.8}{1.7} \\
Cryptic pockets (holo) & 34 & 52.9 & \bmcell{63.7}{2.6} \\
Domain motions & 42 & 56.8 & \bmcell{69.5}{2.5} \\
OOD60 & 38 & 36.8 & \bmcell{45.3}{1.9} \\
Membrane transporters & 30 & 30.0 & \bmcell{24.3}{2.1} \\
Fold switchers & 30 & 40.0 & \bmcell{52.7}{2.6} \\
\bottomrule
\end{tabular}
\end{table*}

\clearpage
\subsection{Results at fixed Pairformer passes}
\label{app:fixed_passes}
In this section, we report the success rates and coverage curves at fixed Pairformer passes for all models.

\begin{table}[h]
\centering
\caption{\textbf{Multi-state benchmarks.} Means and standard deviations of success@100 (using 100 bootstrap trials) are reported. All methods are implemented using OF3p ($\boldsymbol{R=1}$). Ours-coord and Ours-dist denote ConforNets trained with coordinate MSE and distogram CDF MSE objectives, respectively.}
\label{tab:success_r1}
\small
\begin{tabular}{lcccccc}
\toprule
Benchmark & $N$ & OF3p & AFsample3 & ConforMix & Ours-coord & Ours-dist \\
\midrule
Cryptic pockets (apo) & 34 & \bmcell{25.9}{1.2} & \bmcell{39.0}{2.2} & \bmcell{35.8}{2.2} & \bmcell{\textbf{49.7}}{2.8} & \bmcell{48.8}{2.9} \\
Cryptic pockets (holo) & 34 & \bmcell{48.3}{2.8} & \bmcell{65.9}{3.1} & \bmcell{63.2}{2.1} & \bmcell{78.0}{2.6} & \bmcell{\textbf{78.9}}{2.8} \\
Domain motions & 42 & \bmcell{67.5}{2.2} & \bmcell{74.0}{2.3} & \bmcell{74.0}{2.1} & \bmcell{\textbf{81.5}}{1.8} & \bmcell{78.3}{1.8} \\
OOD60 & 38 & \bmcell{34.8}{3.3} & \bmcell{44.6}{3.5} & \bmcell{56.2}{3.5} & \bmcell{53.7}{3.2} & \bmcell{\textbf{60.7}}{2.8} \\
Membrane transporters & 30 & \bmcell{22.3}{2.9} & \bmcell{32.4}{3.1} & \bmcell{34.9}{2.5} & \bmcell{47.2}{3.3} & \bmcell{\textbf{51.1}}{3.7} \\
Fold switchers & 30 & \bmcell{42.8}{2.0} & \bmcell{44.7}{1.6} & \bmcell{49.4}{2.6} & \bmcell{49.9}{1.3} & \bmcell{\textbf{54.4}}{2.4} \\
\bottomrule
\end{tabular}
\end{table}

\begin{figure}[h]
    \centering
    \begin{subfigure}[t]{0.32\linewidth}
        \centering
        \includegraphics[width=\linewidth]{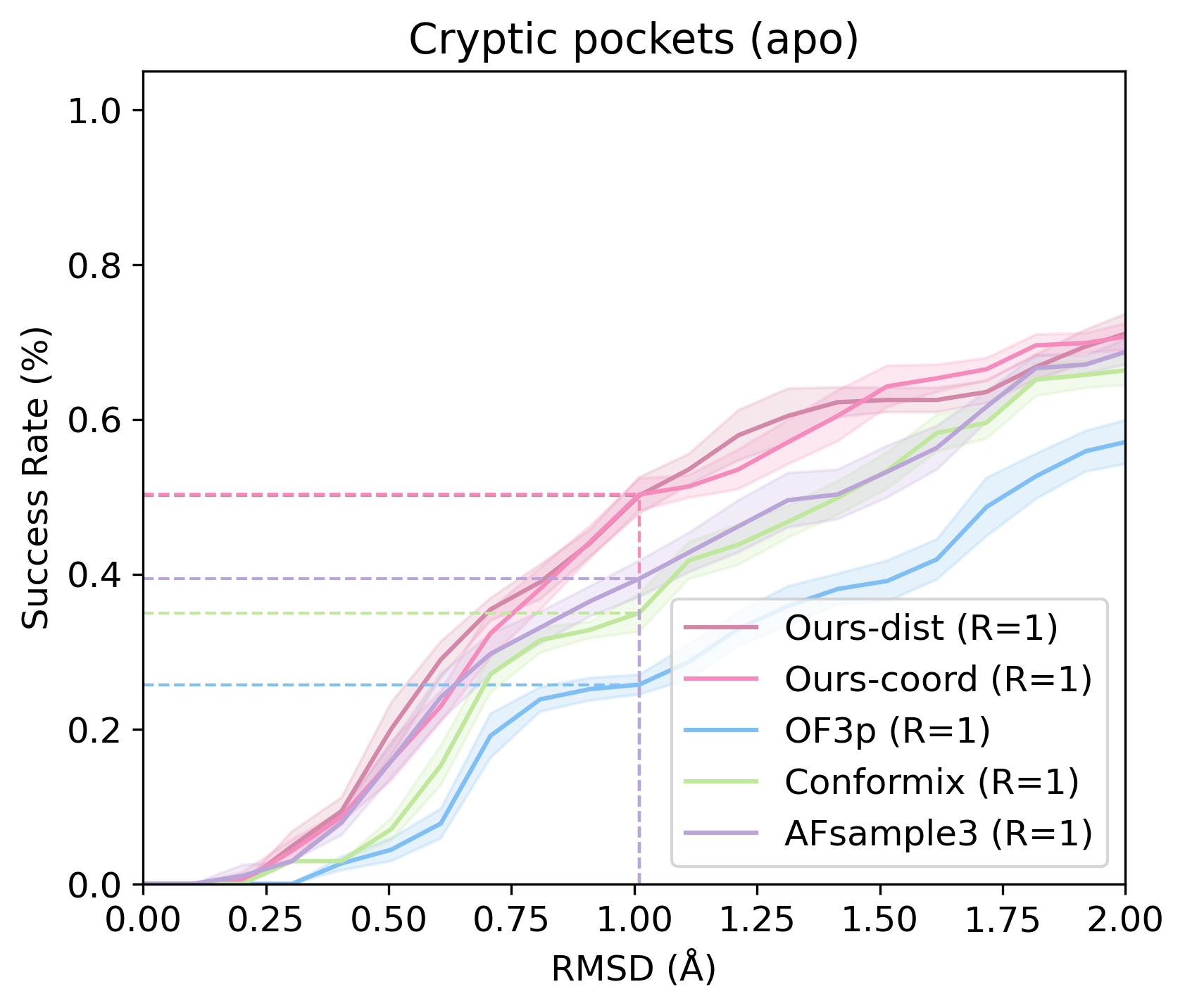}
        \caption{Cryptic pocket (apo)}
    \end{subfigure}\hfill
    \begin{subfigure}[t]{0.32\linewidth}
        \centering
        \includegraphics[width=\linewidth]{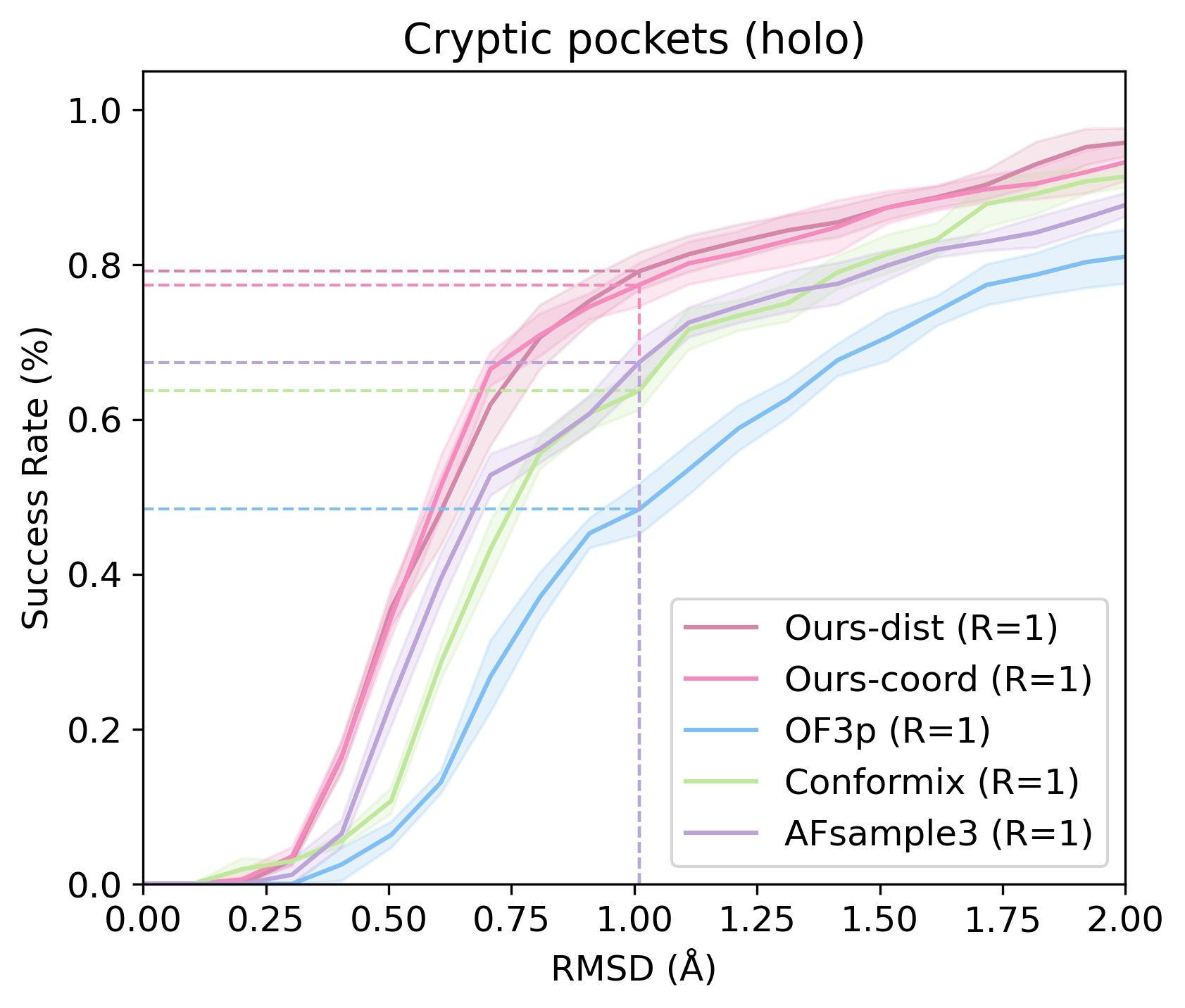}
        \caption{Cryptic pocket (holo)}
    \end{subfigure}\hfill
    \begin{subfigure}[t]{0.32\linewidth}
        \centering
        \includegraphics[width=\linewidth]{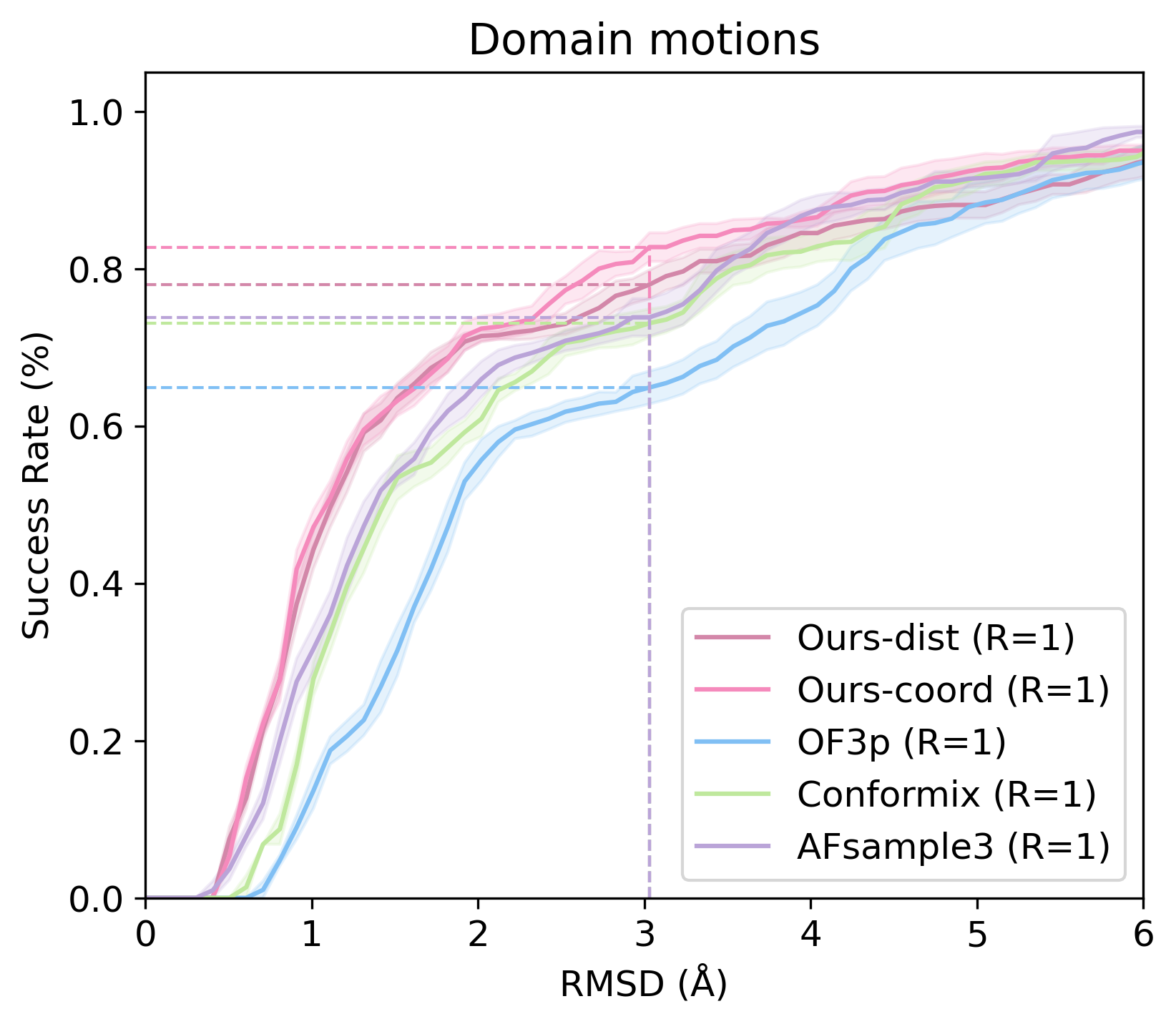}
        \caption{Domain motion}
    \end{subfigure}

    \vspace{0.5em}

    \begin{subfigure}[t]{0.32\linewidth}
        \centering
        \includegraphics[width=\linewidth]{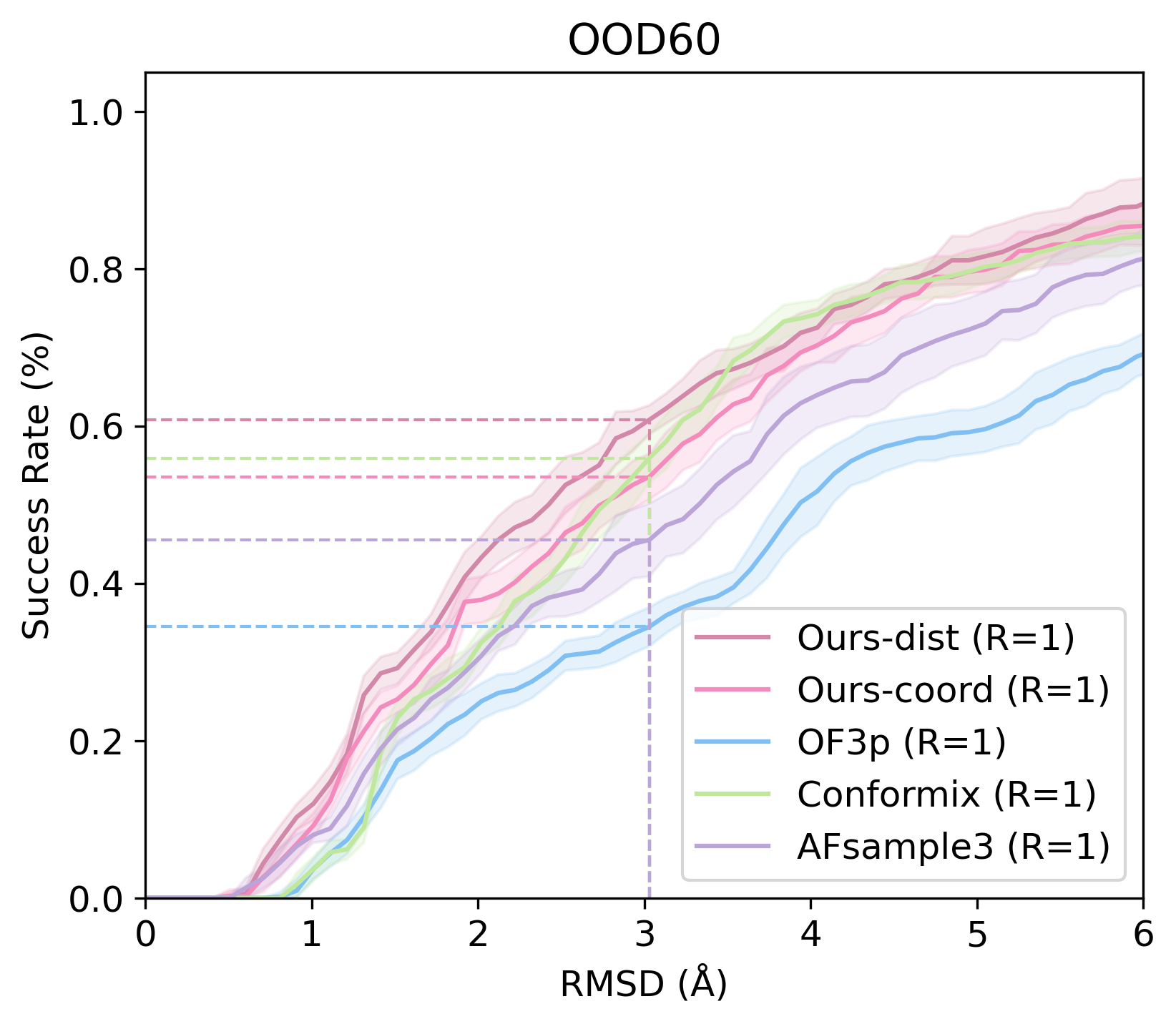}
        \caption{OOD60}
    \end{subfigure}\hfill
    \begin{subfigure}[t]{0.32\linewidth}
        \centering
        \includegraphics[width=\linewidth]{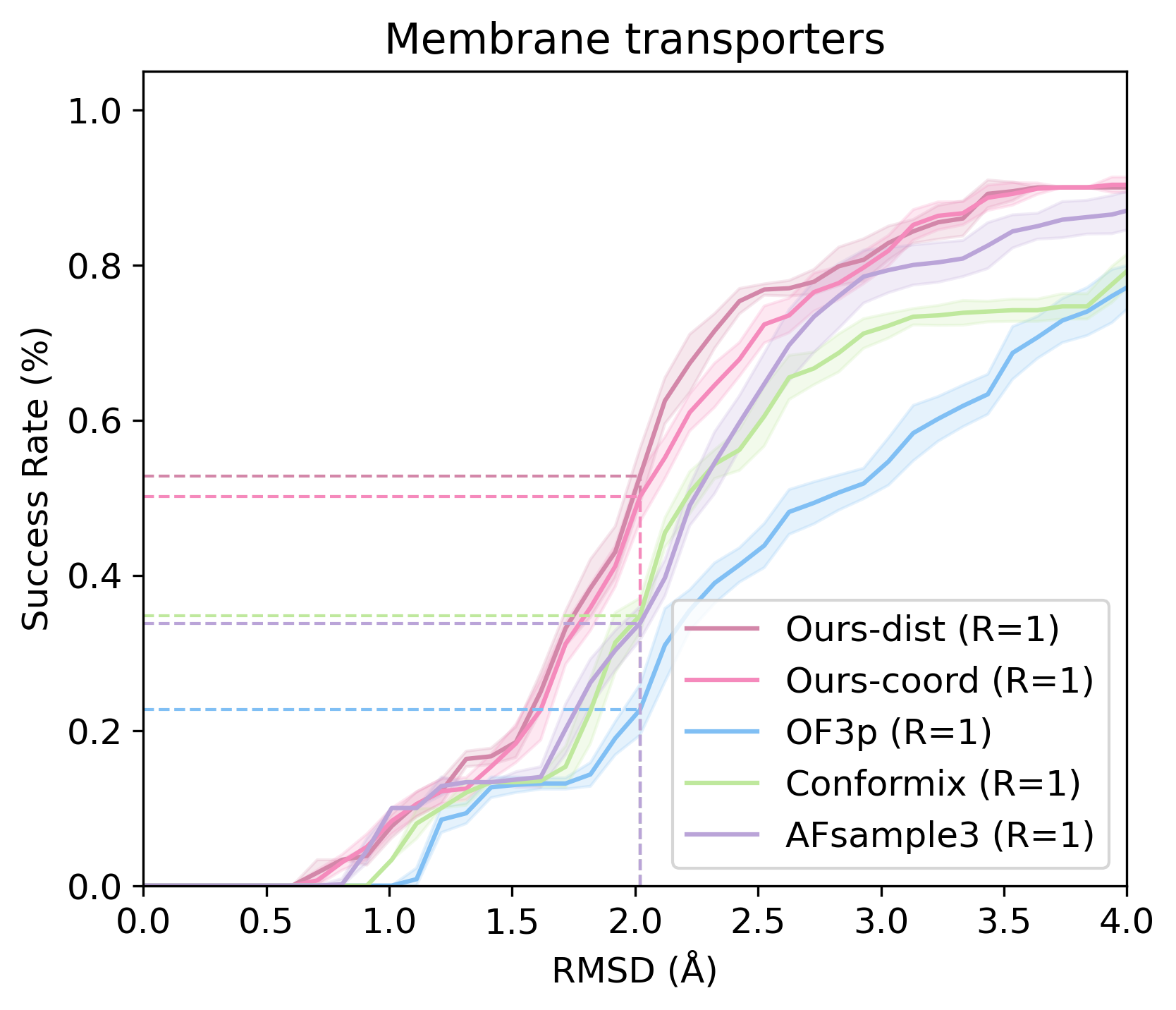}
        \caption{Membrane transporter}
    \end{subfigure}\hfill
    \begin{subfigure}[t]{0.32\linewidth}
        \centering
        \includegraphics[width=\linewidth]{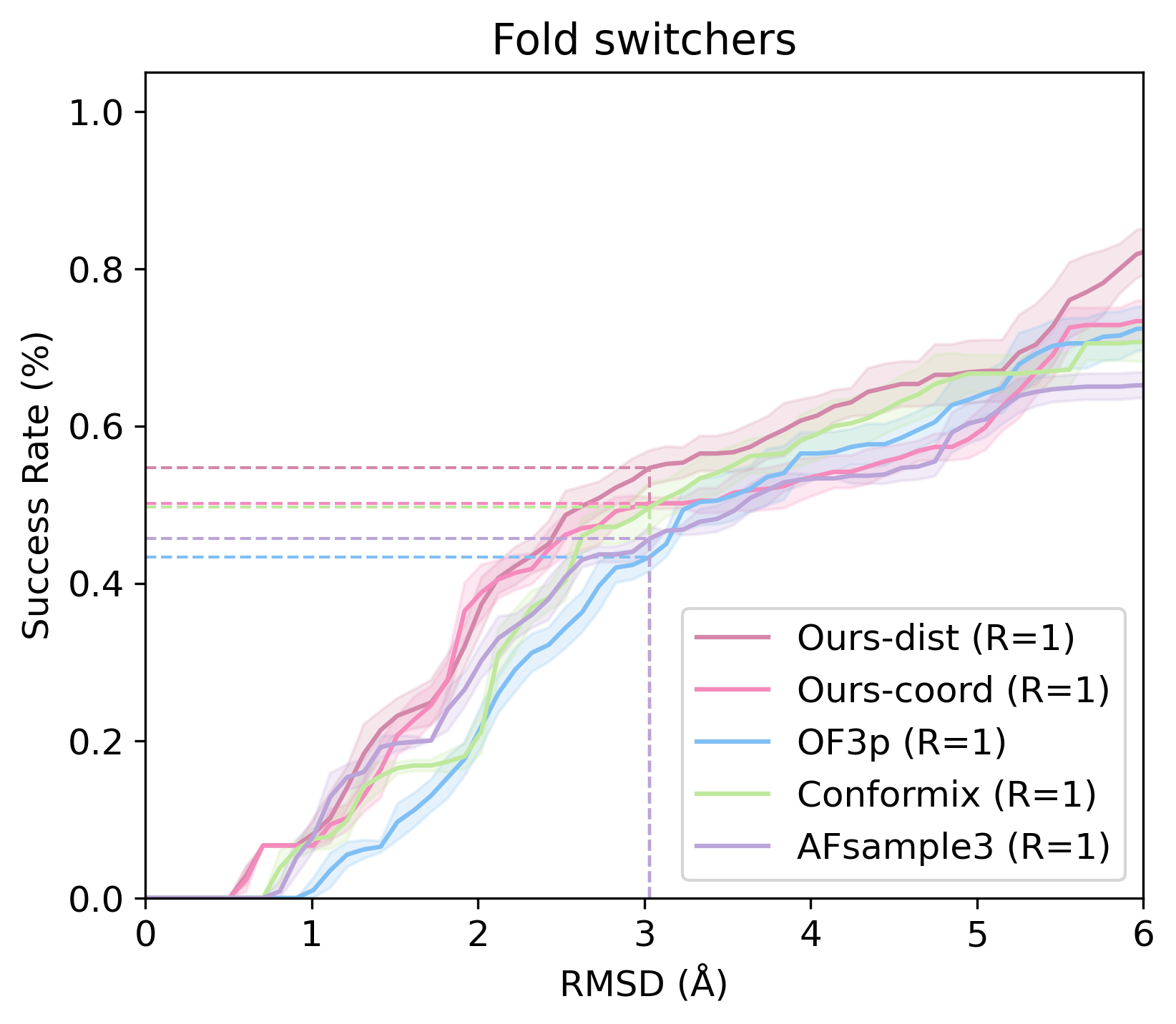}
        \caption{Fold switching}
    \end{subfigure}

    \caption{
        Success@100 rates of covering multi-state benchmarks as a function of RMSD. Shaded regions denote bootstrapped standard deviation. All methods are implemented using OF3p ($\boldsymbol{R=1}$).}
    \label{fig:curve_r1}
\end{figure}

\clearpage

\begin{table}[htbp]
\centering
\caption{\textbf{Multi-state benchmarks.} Means and standard deviations of success@100 (using 100 bootstrap trials) are reported. All methods are implemented using OF3p ($\boldsymbol{R=11}$). Ours-coord and Ours-dist denote ConforNets trained with coordinate MSE and distogram CDF MSE objectives, respectively.}
\small
\label{tab:succes_r11}
\begin{tabular}{lcccccccc}
\toprule
Benchmark & $N$ & OF3p & AFsample3 & ConforMix & Ours-coord & Ours-dist \\
\midrule
Cryptic pockets (apo) & 34 & \bmcell{30.8}{1.7} & \bmcell{44.7}{1.7} & \bmcell{37.0}{1.5} & \bmcell{\textbf{48.2}}{1.8} & \bmcell{46.9}{2.0} \\
Cryptic pockets (holo) & 34 & \bmcell{63.7}{2.6} & \bmcell{73.6}{2.3} & \bmcell{62.7}{2.0} & \bmcell{\textbf{83.0}}{2.5} & \bmcell{76.1}{2.6} \\
Domain motions & 42 & \bmcell{69.5}{2.5} & \bmcell{80.6}{2.3} & \bmcell{80.3}{1.6} & \bmcell{81.7}{1.3} & \bmcell{\textbf{81.9}}{1.4} \\
OOD60 & 38 & \bmcell{45.3}{1.9} & \bmcell{54.0}{3.0} & \bmcell{\textbf{57.7}}{3.3} & \bmcell{51.7}{2.3} & \bmcell{51.5}{2.8} \\
Membrane transporters & 30 & \bmcell{24.3}{2.1} & \bmcell{\textbf{46.9}}{3.4} & \bmcell{34.2}{2.3} & \bmcell{41.8}{3.0} & \bmcell{42.7}{3.4} \\
Fold switchers & 30 & \bmcell{52.7}{2.6} & \bmcell{48.7}{1.8} & \bmcell{\textbf{54.3}}{2.6} & \bmcell{52.5}{2.2} & \bmcell{53.3}{0.3} \\
\bottomrule
\end{tabular}
\end{table}

\begin{figure}[h]
    \centering
    \begin{subfigure}[t]{0.32\linewidth}
        \centering
        \includegraphics[width=\linewidth]{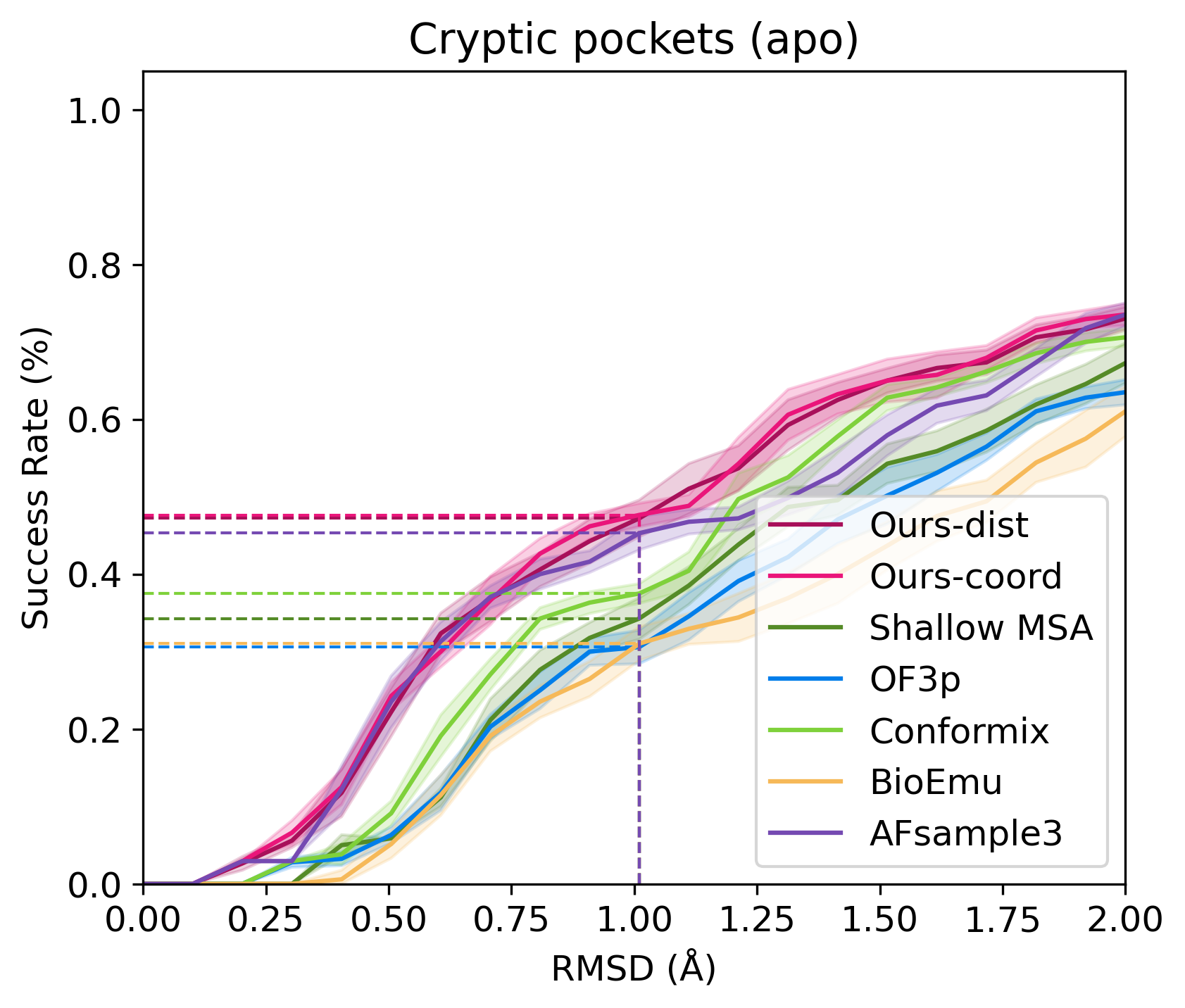}
        \caption{Cryptic pocket (apo)}
    \end{subfigure}\hfill
    \begin{subfigure}[t]{0.32\linewidth}
        \centering
        \includegraphics[width=\linewidth]{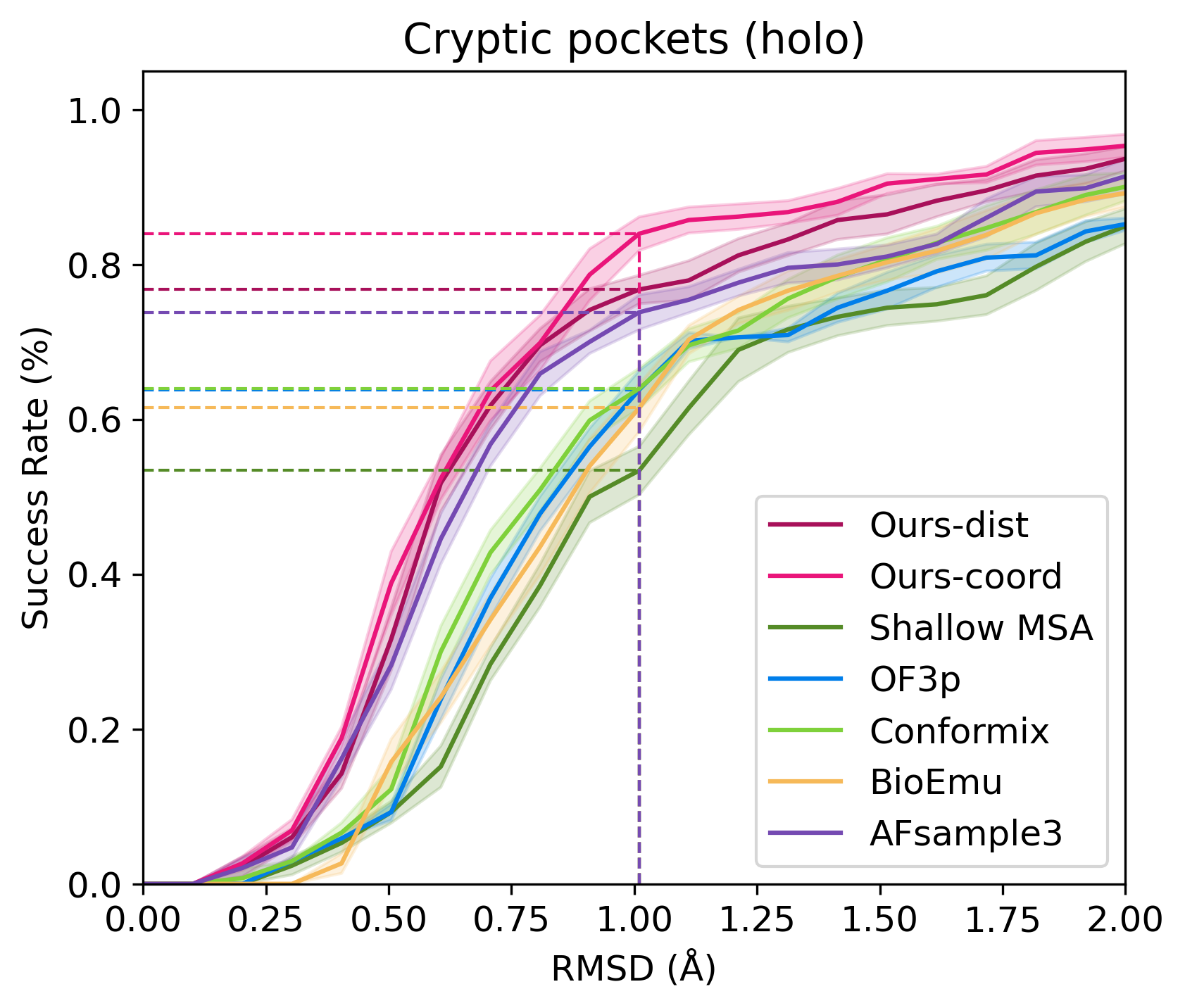}
        \caption{Cryptic pocket (holo)}
    \end{subfigure}\hfill
    \begin{subfigure}[t]{0.32\linewidth}
        \centering
        \includegraphics[width=\linewidth]{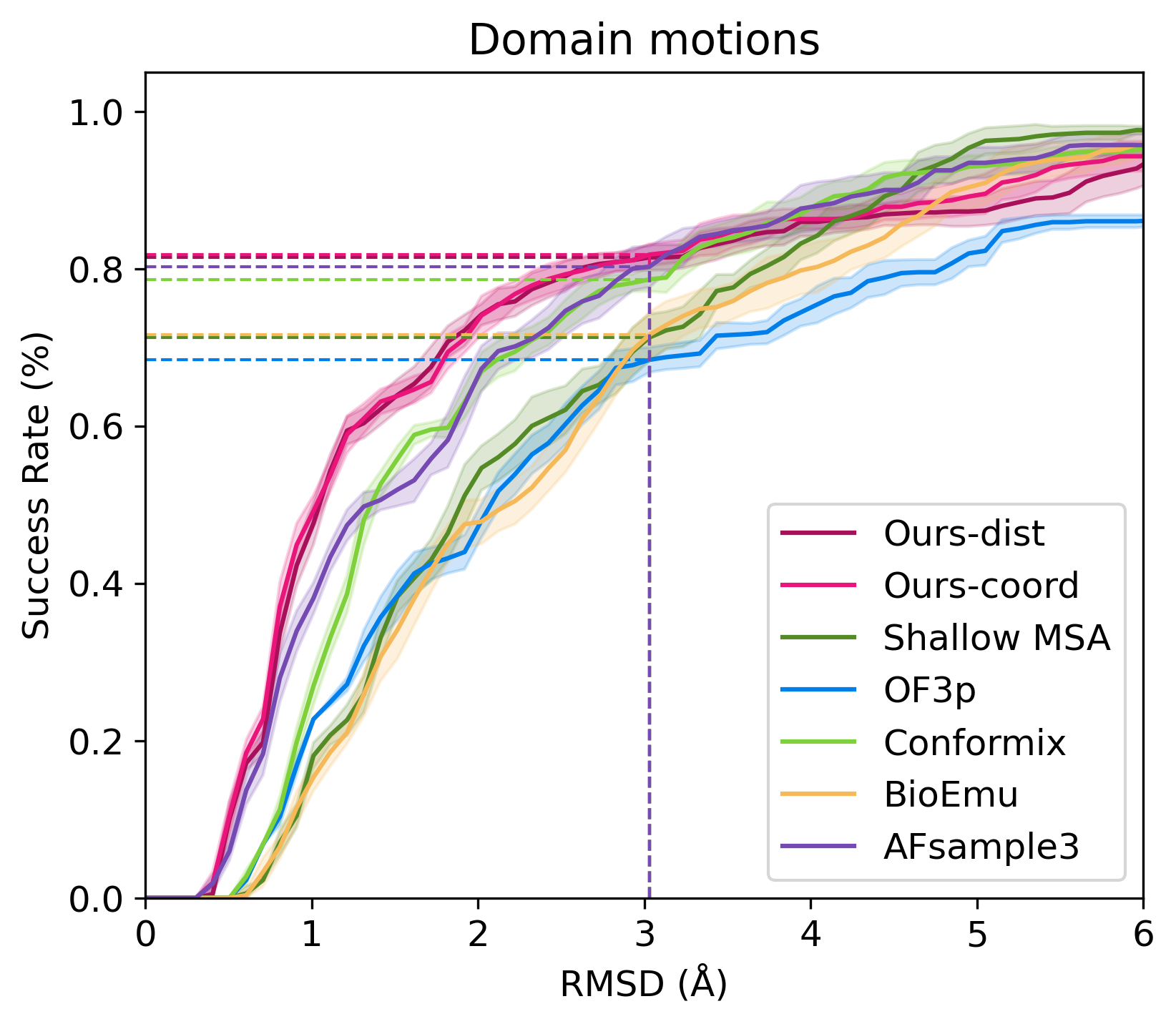}
        \caption{Domain motion}
    \end{subfigure}

    \vspace{0.5em}

    \begin{subfigure}[t]{0.32\linewidth}
        \centering
        \includegraphics[width=\linewidth]{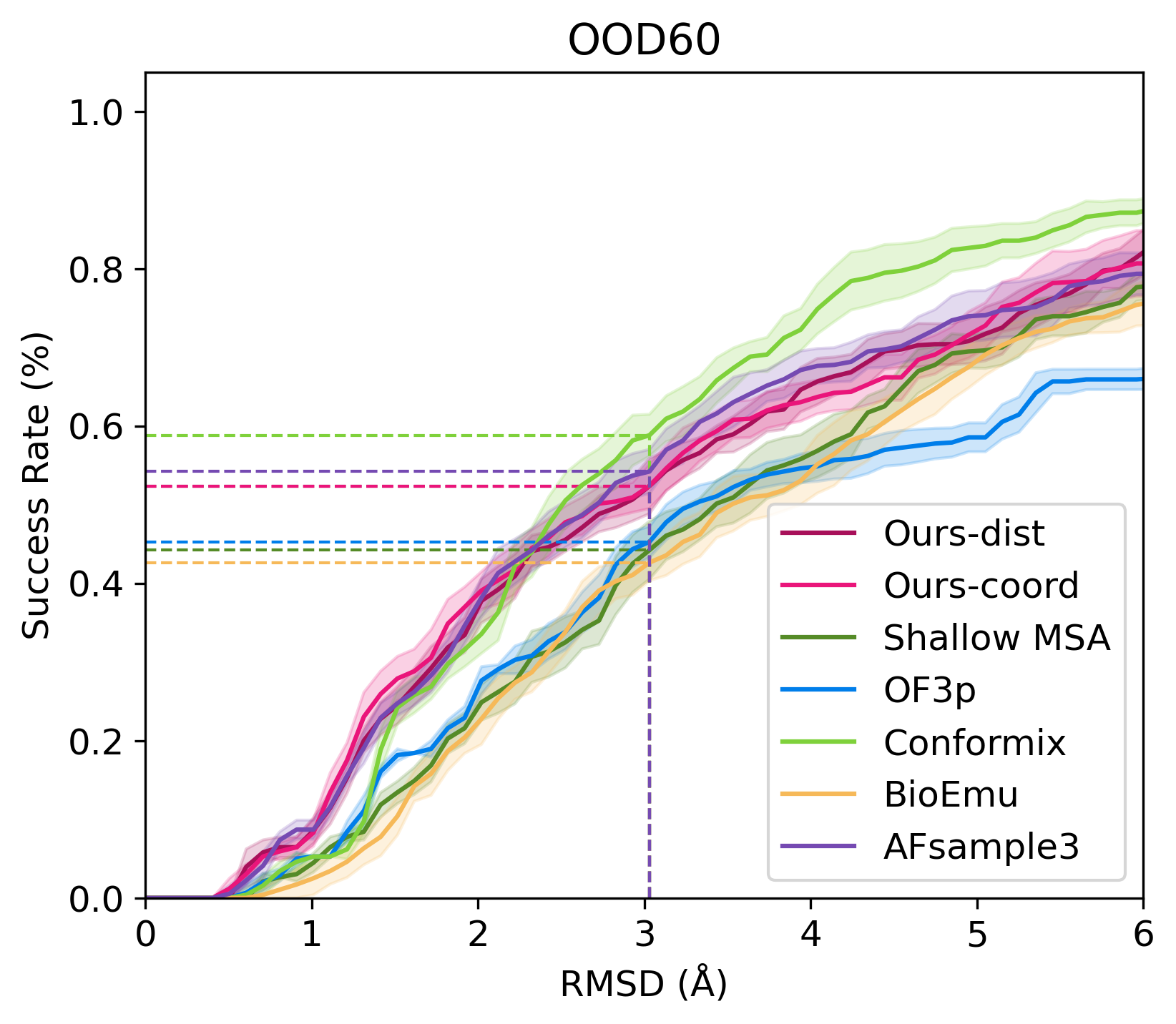}
        \caption{OOD60}
    \end{subfigure}\hfill
    \begin{subfigure}[t]{0.32\linewidth}
        \centering
        \includegraphics[width=\linewidth]{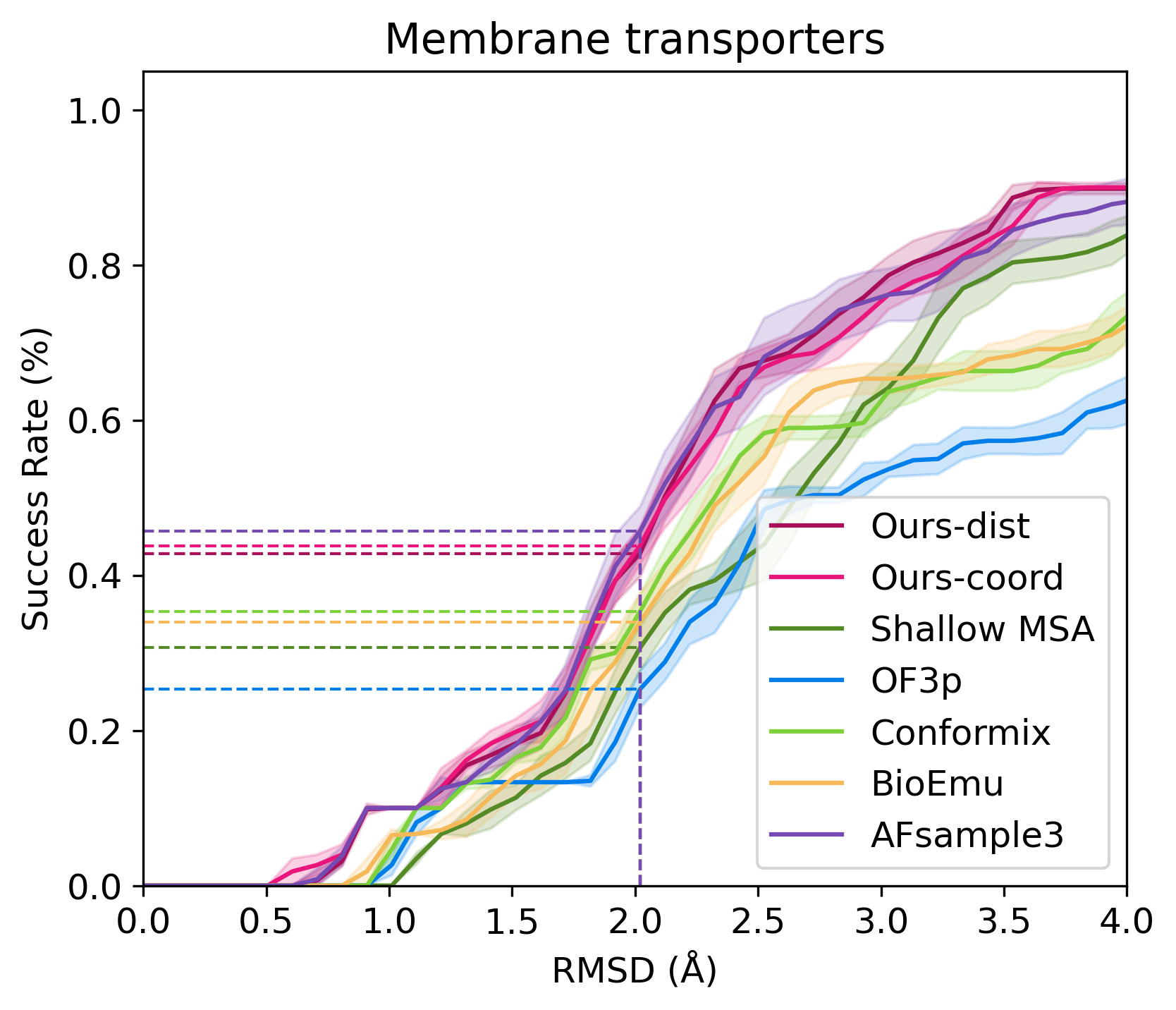}
        \caption{Membrane transporter}
    \end{subfigure}\hfill
    \begin{subfigure}[t]{0.32\linewidth}
        \centering
        \includegraphics[width=\linewidth]{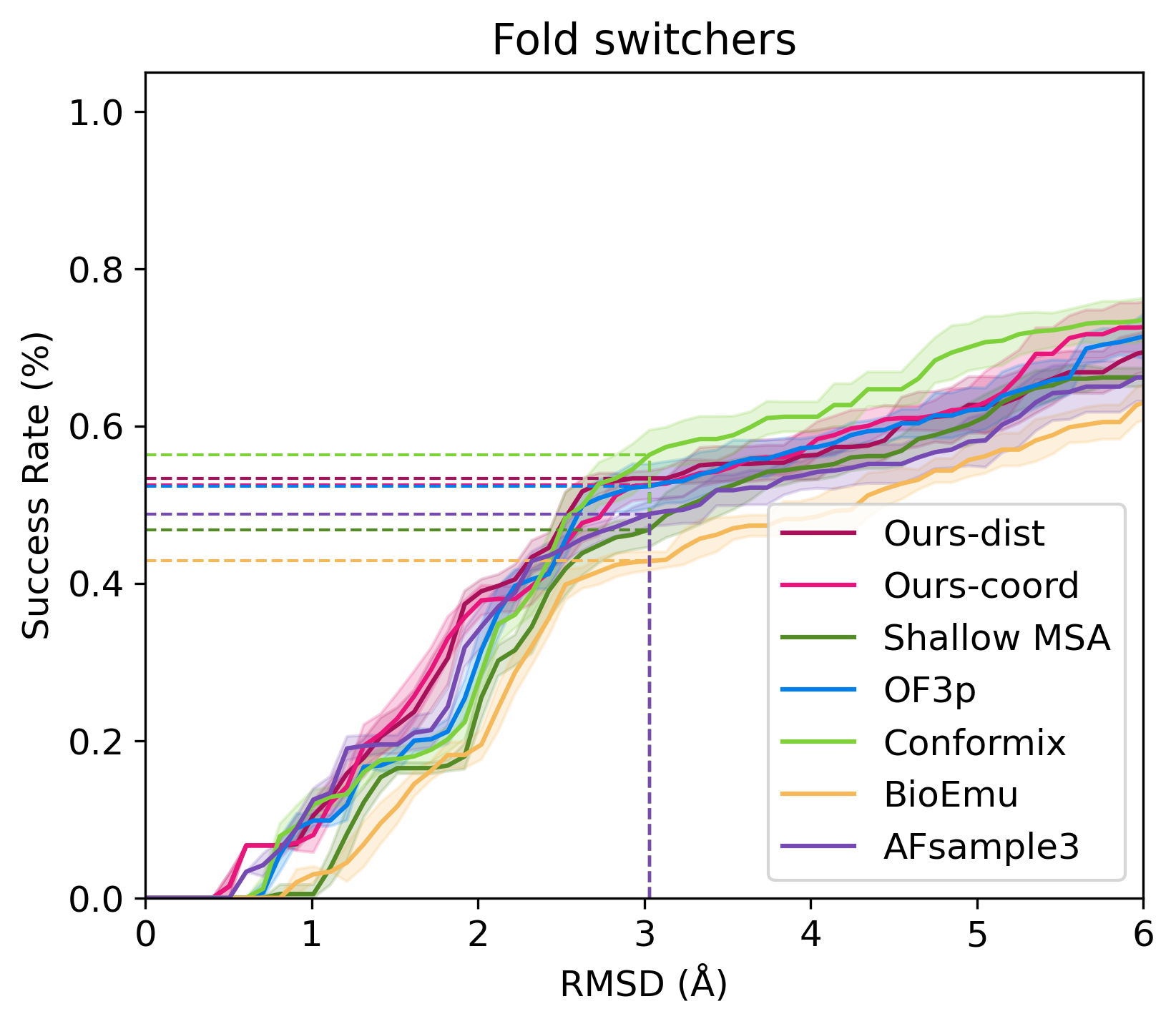}
        \caption{Fold switching}
    \end{subfigure}

    \caption{
    Success@100 rates of covering multi-state benchmarks as a function of RMSD. Shaded regions denote bootstrapped standard deviation. All methods except BioEmu are implemented using OF3p ($\boldsymbol{R=11}$).}
    \label{fig:curve_r11}
\end{figure}

\clearpage 
\subsection{Providing ligand as input in cryptic pocket benchmarks}
\label{app:holo_query}

As shown in Table~\ref{tab:holo_query_success}, providing the ligand input to OF3p improves holo conformation prediction on the cryptic pocket benchmark. 

\begin{table*}[h]
\centering
\caption{\textbf{Cryptic pockets benchmark performance with and without ligand input to OF3p ($\boldsymbol{R=1}$).} Means and standard deviations of success@100 (using 100 bootstrap trials) are reported.}
\small
\label{tab:holo_query_success}
\begin{tabular}{lccc}
\toprule
Conformation & $N$ & OF3p (R=1) with ligand & OF3p (R=1) without ligand \\
\midrule
Apo & 34 & \bmcell{25.4}{2.0} & \bmcell{26.0}{1.1} \\
Holo & 34 & \bmcell{63.7}{3.1} & \bmcell{47.9}{2.5} \\
\bottomrule
\end{tabular}
\end{table*}

\begin{figure}[h]
    \centering
    \begin{subfigure}[t]{0.32\linewidth}
        \centering
        \includegraphics[width=\linewidth]{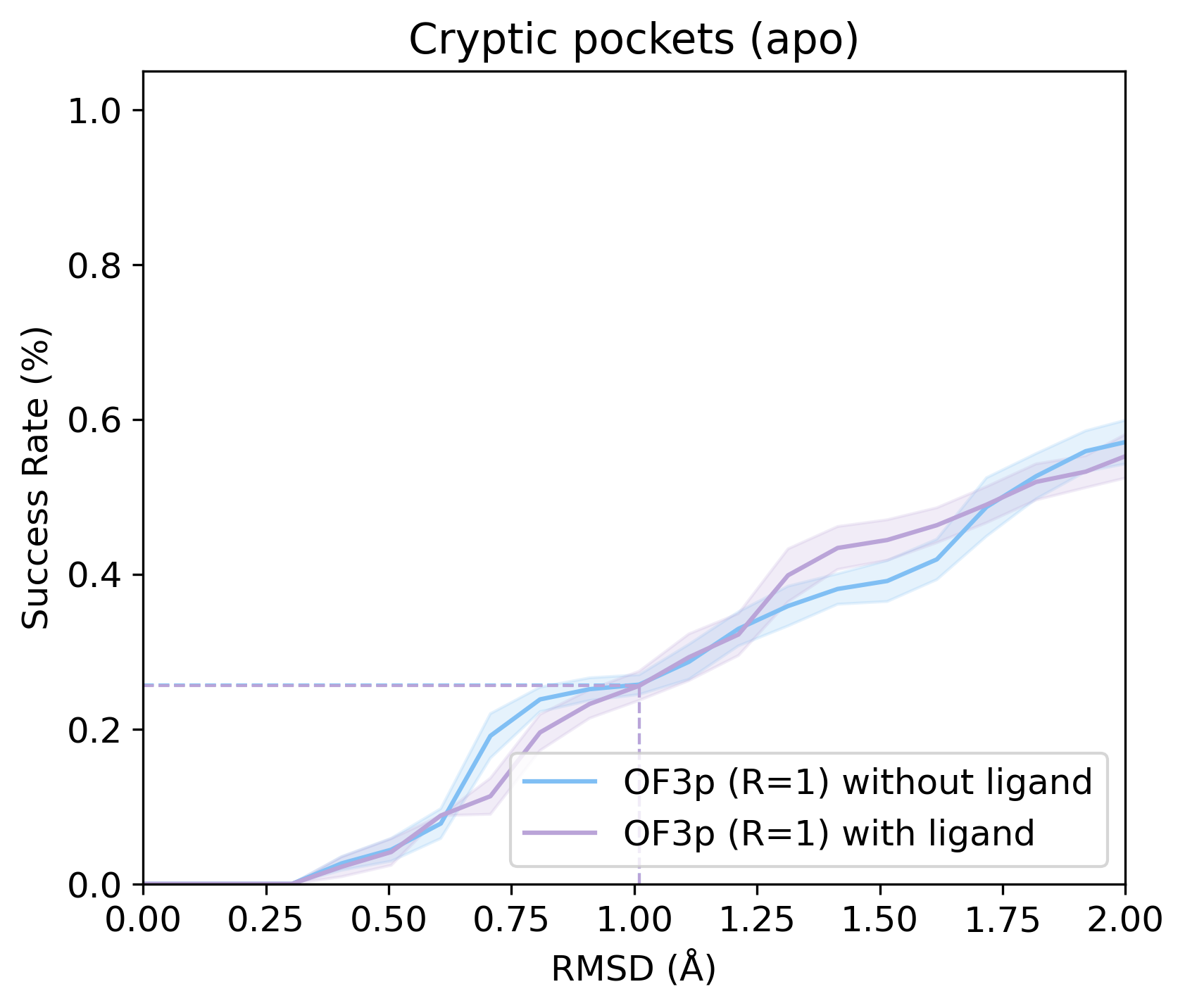}
        \caption{Apo}
    \end{subfigure}
    \begin{subfigure}[t]{0.32\linewidth}
        \centering
        \includegraphics[width=\linewidth]{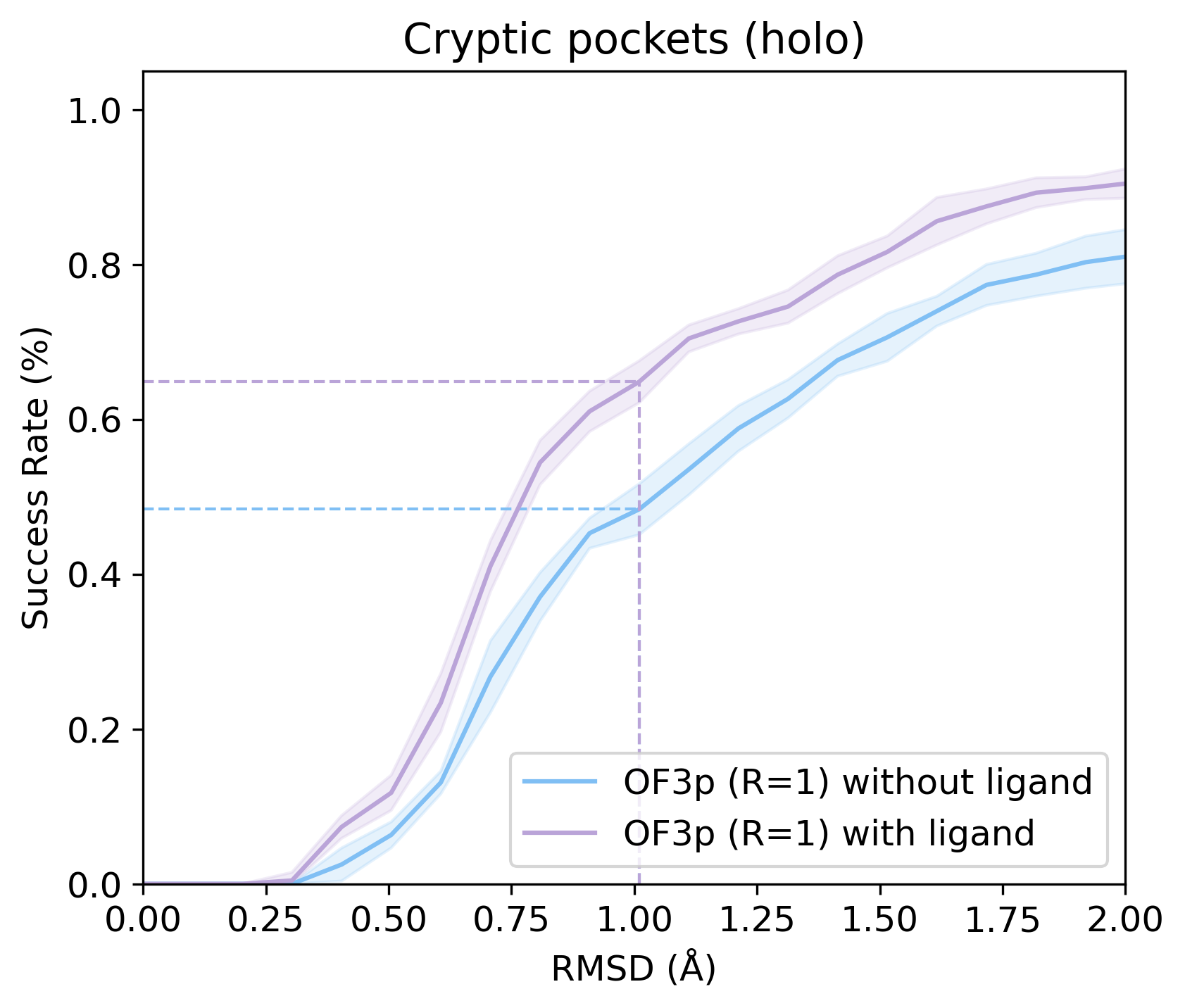}
        \caption{Holo}
    \end{subfigure}
    \caption{Success@100 rates of covering cryptic pockets in the holo and apo states as a function of RMSD.}
\end{figure}

\subsection{Ablation on diversity objective}
\label{app:objective_ablation}

In this section, we compare three diversity training objectives that differ in where diversity is enforced: coordinate space,  distogram space, and pair representation space, along with an entropy maximization baseline. As shown in Table~\ref{tab:ablation}, all diversity objectives perform comparably and substantially outperform the entropy maximization baseline, indicating that explicitly encouraging diversity is more effective than using distogram entropy as a proxy. The coordinate-based objective performs best on the cryptic pocket benchmark. The distogram- and pair-space objectives remain competitive overall, suggesting that diversity can often be enforced without materializing a structure through a diffusion model.

\begin{table*}[h]
\centering
\caption{\textbf{Multi-state benchmarks performance across diversity objectives.} Means and standard deviations of success@100 (using 100 bootstrap trials) are reported. All methods are implemented using OF3p ($\boldsymbol{R=1}$).}
\small
\label{tab:ablation}
\begin{tabular}{lccccc}
\toprule
Benchmark & $N$ & Distogram CDF MSE & Coordinate MSE & Pair representation MSE & Entropy maximization \\
\midrule
Cryptic pockets (apo) & 34 & \bmcell{49.1}{2.9} & \bmcell{49.0}{2.8} & \bmcell{47.4}{2.4} & \bmcell{37.8}{3.5} \\
Cryptic pockets (holo) & 34 & \bmcell{79.4}{2.4} & \bmcell{77.5}{2.4} & \bmcell{76.5}{2.8} & \bmcell{70.1}{3.7} \\
Domain motions & 42 & \bmcell{77.8}{1.6} & \bmcell{81.5}{1.9} & \bmcell{80.0}{2.1} & \bmcell{72.6}{2.0} \\
OOD60 & 38 & \bmcell{60.4}{3.1} & \bmcell{53.9}{3.6} & \bmcell{59.4}{3.9} & \bmcell{46.9}{3.4} \\
Membrane transporters & 30 & \bmcell{50.4}{3.9} & \bmcell{47.0}{3.4} & \bmcell{48.0}{4.1} & \bmcell{38.3}{3.5} \\
Fold switchers & 30 & \bmcell{54.1}{2.7} & \bmcell{50.0}{1.7} & \bmcell{54.8}{2.1} & \bmcell{47.1}{2.9} \\
\bottomrule
\end{tabular}
\end{table*}

\clearpage 

\section{Additional results on conformation transfer}

\subsection{Source selection}

In this section, we analyze how transfer performance depends on the choice of source protein.

\paragraph{Experimental design}
For GPCRs, we randomly selected 8 proteins as sources, while for kinases and transporters, we evaluated all proteins as sources. We trained $N$ ConforNets for each source protein and applied each of them to all other proteins in the benchmark, generating 10 diffusion samples per transfer. We use $N=10$ for GPCRs, $N=8$ for membrane transporters, and $N=4$ for kinases, with smaller $N$ for some benchmarks due to compute constraints. 

\paragraph{Metrics} In addition to success@5, we also quantify the expected minimum RMSD over 5 samples to provide a more fine-grained view. For each ConforNet, we draw 5 samples without replacement from the pool of 10 diffusion samples, compute the minimum RMSD, repeat this procedure over 100 bootstrap trials, and then average across trials and across the $N$ ConforNets trained for that source.

\begin{figure}[h!]
    \centering
    \includegraphics[width=\linewidth]{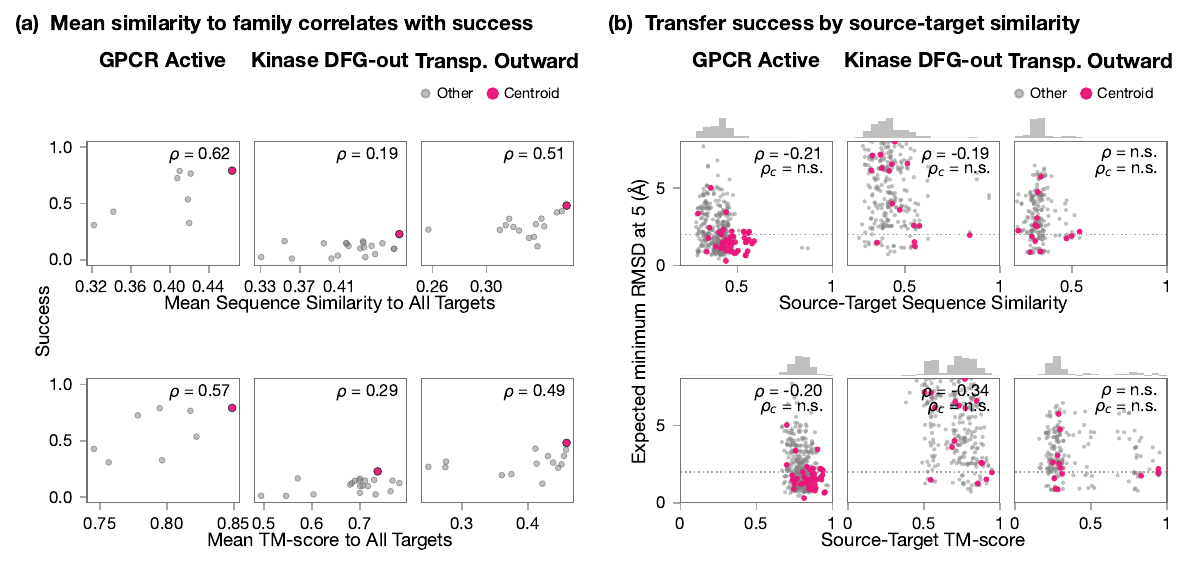}
    \caption{\textbf{(a)} Success@5 by mean sequence similarity and mean TM-score to other benchmark targets (Spearman $\rho$ shown).
    \textbf{(b)} RMSD between prediction and target is plotted against source-target sequence similarity and TM-score, with the centroid source highlighted pink. Spearman correlation coefficients for all sources (\(\rho\)) and for centroid sources (\(\rho_C\)) are shown. n.s. denotes not significant (\(p > 0.05\)).}
    \label{fig:transfer_similarity}
\end{figure}

\paragraph{The centroid protein is the best-performing source.} 
In Figure~\ref{fig:transfer_similarity}a, we observe a strong correlation between mean sequence similarity to all other members of the family and success@5 for GPCRs and transporters. For all three benchmarks, the centroid protein is the best-performing source.

\paragraph{Transfer success depends only weakly on source-target similarity.} Figure~\ref{fig:transfer_similarity}b visualizes transfer performance for all source-target pairs. Across all source-target pairs, we observe at most a weak negative correlation. When the analysis is restricted to the centroid source, this trend disappears ($\rho_C$ is not significant in all benchmarks). This weak dependence has two implications. First, successful transfers with sub-2\AA{} RMSD are achieved even at low similarity, as seen in the lower-left region of each panel. Second, because transfer success is only weakly predicted by similarity, source selection need not be tailored to each target. In practice, choosing a source that is simply central within the family is sufficient. Together, these results suggest that the ConforNet captures a conformational mode that is orthogonal to coarse sequence or fold similarity, and they justify the practical strategy of training once on the centroid source and transferring across the family.

\label{app:source_selection}

\subsection{Effect of Pairformer recycles in predicting rare conformation}
\label{app:recycle_transfer_baseline}

In the transfer setting, increasing the number of Pairformer passes from $R=1$ to $R=11$ did not improve performance for baselines (Table~\ref{tab:transfer_success_5_100_recycles}). Reachability drops in 4 out of 6 benchmark/method combinations, despite using much more compute. One possible reason is that our targets are monomeric proteins, for which even a single pass is often sufficient to produce structurally reasonable predictions. Also, our goal is to induce a rare conformational mode. Because the MSA is resampled at each recycle, additional recycles may repeatedly nudge the model toward the dominant conformation preferred by the MSA. Based on these results, and due to compute constraints, we use $R=1$ in all transfer experiments in the main text. It would still be interesting in future work to test ConforNets transfer with $R=11$ because we intervene directly on pair representations rather than through MSA perturbations.

\begin{table*}[th!]
\centering
\caption{\textbf{Conformation transfer performance of baselines across Pairformer passes}, evaluated by reachability and at-will induction for three benchmark protein families. We only generate 100 samples per benchmark/model combination and therefore do not report standard deviation from bootstrap trials for success@100.} 
\label{tab:transfer_success_5_100_recycles}
\small
\begin{tabular}{lcccccc}
\toprule
& \multicolumn{3}{c}{\textbf{At-will Induction} (success@5)} & \multicolumn{3}{c}{\textbf{Reachability} (success@100)} \\
\cmidrule(lr){2-4} \cmidrule(lr){5-7}
Method & GPCR active & Kinase DFG-out & Transp. Out & GPCR active & Kinase DFG-out & Transp. Out \\
\midrule
OF3p ($R=1$)
& \bmcell{24.2}{2.4}
& \bmcell{5.8}{2.3}
& \bmcell{15.6}{5.2}
& 37.3
& 10.0
& 33.3 \\
OF3p ($R=11$)
& \bmcell{9.1}{2.1}
& \bmcell{14.4}{3.3}
& \bmcell{11.9}{4.6}
& 15.7
& 20.0
& 20.0 \\
AFsample3 ($R=1$)
& \bmcell{27.7}{4.2}
& \bmcell{4.4}{3.7}
& \bmcell{20.4}{5.3}
& 60.8
& 20.0
& 33.3 \\
AFsample3 ($R=11$)
& \bmcell{26.1}{4.5}
& \bmcell{5.9}{2.3}
& \bmcell{25.6}{5.1}
& 51.0
& 15.0
& 33.3 \\
\bottomrule
\end{tabular}
\end{table*}

\clearpage

\section{Extended results}

\subsection{Optimal perturbation location}
This section provides the full results corresponding to Sec.~\ref{sec:results_perturbation}. In the main text, we report the $K=1$ setting as a representative comparison across perturbation locations. Here, we additionally show the full sweep over mini-rollout length $K \in \{1,2,5,10\}$. Table~\ref{tab:gt_sweep_full} shows consistent trends across all $K$: perturbing the pre-Pairformer pair latent ($\mathbf{z}^{\mathrm{pre}}$) achieve low RMSD under both mini and full rollouts, indicating stable control that persists under standard sampling. In contrast, perturbing the post-Pairformer latents can fit the mini rollout but degrade under full diffusion, most notably for $\mathbf{s}^{\mathrm{post}}$, suggesting shortcut solutions that do not survive longer denoising trajectories. Adapting $\mathbf{s}^{\mathrm{pre}}$ performs poorly for all $K$, suggesting that the gap is not primarily due to the mini-rollout approximation, but rather reflects limited control from $\mathbf{s}^{\mathrm{pre}}$ itself.

\begin{table}[h]
\centering
\caption{Mean and standard deviation of RMSD between the ConforNet-induced predictions and the ground truth entries of OOD60 using different perturbation locations and mini-rollout length $K$. Statistics are computed across 38 entries $\times$ 4 replicates. ConforNets are trained directly to ground truths using $K$ mini rollouts.}
\label{tab:gt_sweep_full}
\small
\setlength{\tabcolsep}{4pt}
\begin{subtable}{.4\linewidth}
\centering
\caption{Pair representations}
\begin{tabular}{lccc}
\toprule
Placement & $K$ & Mini & Full \\
\midrule
$\mathbf{z}^{\mathrm{post}}$ & 1  & 1.79 $\pm$ 1.19 & 3.40 $\pm$ 2.37 \\
$\mathbf{z}^{\mathrm{post}}$ & 2  & 1.61 $\pm$ 1.11 & 2.41 $\pm$ 1.74 \\
$\mathbf{z}^{\mathrm{post}}$ & 5  & 1.65 $\pm$ 1.43 & 2.38 $\pm$ 2.08 \\
$\mathbf{z}^{\mathrm{post}}$ & 10 & 1.64 $\pm$ 1.05 & 2.23 $\pm$ 1.92 \\
\midrule
$\mathbf{z}^{\mathrm{pre}}$  & 1  & 1.90 $\pm$ 2.05 & 1.93 $\pm$ 1.55 \\
$\mathbf{z}^{\mathrm{pre}}$  & 2  & 1.72 $\pm$ 1.13 & 1.81 $\pm$ 1.11 \\
$\mathbf{z}^{\mathrm{pre}}$  & 5  & 2.10 $\pm$ 1.77 & 1.94 $\pm$ 1.41 \\
$\mathbf{z}^{\mathrm{pre}}$  & 10 & 2.46 $\pm$ 2.49 & 2.08 $\pm$ 1.40 \\
\bottomrule
\end{tabular}
\end{subtable}
\begin{subtable}{.4\linewidth}
\centering
\caption{Single representations}
\begin{tabular}{lccc}
\toprule
Placement & $K$ & Mini & Full \\
\midrule
$\mathbf{s}^{\mathrm{post}}$ & 1  & 2.31 $\pm$ 2.99 & 9.84 $\pm$ 26.33 \\
$\mathbf{s}^{\mathrm{post}}$ & 2  & 2.19 $\pm$ 4.00 & 7.09 $\pm$ 27.19 \\
$\mathbf{s}^{\mathrm{post}}$ & 5  & 1.99 $\pm$ 3.17 & 6.81 $\pm$ 27.68 \\
$\mathbf{s}^{\mathrm{post}}$ & 10 & 2.09 $\pm$ 3.30 & 4.48 $\pm$ 9.19 \\
\midrule
$\mathbf{s}^{\mathrm{pre}}$  & 1  & 4.14 $\pm$ 3.78 & 4.41 $\pm$ 3.85 \\
$\mathbf{s}^{\mathrm{pre}}$  & 2  & 4.37 $\pm$ 4.08 & 4.36 $\pm$ 3.85 \\
$\mathbf{s}^{\mathrm{pre}}$  & 5  & 4.42 $\pm$ 4.51 & 4.50 $\pm$ 4.37 \\
$\mathbf{s}^{\mathrm{pre}}$  & 10 & 4.47 $\pm$ 4.04 & 4.34 $\pm$ 4.03 \\
\bottomrule
\end{tabular}
\end{subtable}
\end{table}

\subsection{Compute overhead}
\label{app:compute_overhead}

This section supplements Sec.~\ref{sec:discussion_time}. Comparing compute across methods is not straightforward, since the cost depends on several implementation choices, including the number of Pairformer passes $R$, the number of diffusion samples, and the diffusion batch size. For ConforMix, the extra cost is incurred during diffusion sampling because it uses Twisted Diffusion Sampling \citep{wu2023practical}, and therefore scales with the number of diffusion forward passes. For ConforNets, in contrast, the extra cost is paid once during training, and can then be amortized over subsequent samples. Applying a trained ConforNet at inference time adds negligible overhead, since it is just an affine transformation.

We therefore report wall-clock runtimes only in a fixed setting matching our experiments: 5 diffusion samples, diffusion batch size 5 (thus a single diffusion forward pass), and either $R=11$ or $R=1$ Pairformer passes. Table~\ref{tab:wallclock_time} compares trained ConforNet sampling, default OF3p sampling, and ConforMix in this setting. ConforNets runtime includes both the amortized training cost and the sampling cost. ConforMix runtime does not includes inference time for baseline prediction. In this regime, ConforNets cost roughly $2$--$3\times$ default OF3p sampling, comparable to ConforMix.

\begin{table}[h]
\centering
\caption{Wall-clock runtime comparison in seconds, reported per 5 diffusion samples on a single 40GB A100 GPU. ConforNets runtime includes amortized training cost and sampling cost.}
\label{tab:wallclock_time}
\small
\setlength{\tabcolsep}{6pt}
\begin{subtable}{.45\linewidth}
\centering
\caption{$R=11$}
\begin{tabular}{lccc}
\toprule
Length & ConforNets & OF3p & ConforMix \\
\midrule
300 & 37 & 19 & 42 \\
400 & 64 & 26 & 54 \\
\bottomrule
\end{tabular}
\end{subtable}
\begin{subtable}{.45\linewidth}
\centering
\caption{$R=1$}
\begin{tabular}{lccc}
\toprule
Length & ConforNets & OF3p & ConforMix \\
\midrule
300 & 31 & 13 & 35 \\
400 & 51 & 16 & 40 \\
\bottomrule
\end{tabular}
\end{subtable}
\end{table}

To contextualize the training-side cost of ConforNets, we also report the runtime and peak memory usage of ConforNet optimization itself in Table~\ref{tab:time_memory}. These measurements correspond only to the 20 Pairformer backpropagation steps used during training, with gradient checkpointing applied at each Pairformer block. As expected, training cost grows with sequence length, and training requires additional memory relative to inference-time methods.

\begin{table}[h!]
\centering
\caption{Runtime and peak memory usage of 20 Pairformer backpropagation steps at different sequence lengths. Measured on a single 80GB A100 GPU.}
\label{tab:time_memory}
\small
\begin{tabular}{rcc}
\toprule
Length & Time (s) & Memory (GB) \\
\midrule
100 & 25.89 & 2.66 \\
200 & 38.63 & 5.21 \\
300 & 90.04 & 9.84 \\
400 & 190.01 & 17.04 \\
500 & 366.11 & 26.85 \\
\bottomrule
\end{tabular}
\end{table}

Overall, both ConforNets and ConforMix introduce moderate additional wall-clock cost relative to default OF3p sampling in our experimental regime. 

\subsection{Analysis of learned ConforNets}
\label{app:weights_extended}
\begin{figure}[ht!]
    \centering
    \includegraphics[width=.6\linewidth]{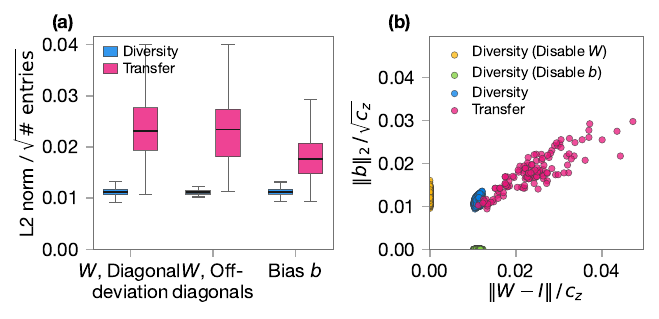}
    \caption{\textbf{Magnitude of learned ConforNet parameters $(\mathbf{W}, \mathbf{b})$.} \textbf{(a)} Distribution of per-ConforNet L2 norms of the diagonal deviation $\mathrm{diag}(\mathbf{W}-\mathbf{I})$, the off-diagonal entries of $\mathbf{W}$, and the bias $\mathbf{b}$, each normalized by the number of entries, comparing diversity (coordinate MSE objective) and transfer runs. \textbf{(b)} Per-ConforNet L2 norms of the weight deviation from identity $\mathbf{W}-\mathbf{I}$ and the bias $\mathbf{b}$ across diversity ablations and transfer runs, normalized by $c_z$ and $\sqrt{c_z}$, respectively.}

    \label{fig:wb_weights}
\end{figure}

This section supplements Sec.~\ref{sec:analysis_weights}.
Figure~\ref{fig:wb_weights} compares the magnitudes of learned $\mathbf{W}$ and $\mathbf{b}$ across diversity and transfer runs, showing that transfer-trained ConforNets have both larger parameter norms and broader spreads. 

Table~\ref{tab:membrane_affine_ablation} shows that removing $\mathbf{W}$ or restricting it to be diagonal reduces performance, indicating that cross-channel mixing contributes meaningfully despite the small absolute magnitude of off-diagonal terms. 

\begin{table}[h]
\centering
\small
\caption{\textbf{Membrane transporters benchmark performance across affine ablations.} Means and standard deviations of success @100 (using 100 bootstrap trials) are reported. All ConforNets are trained with the coordinate MSE objective. We use $R=1$, and the number of sampling seeds follows Sec.~\ref{sec:baselines}.}
\label{tab:membrane_affine_ablation}
\begin{tabular}{lc}
\toprule
Ablation & Success rate \\
\midrule
Default (rerun) & \bmcell{47.5}{3.7} \\
Disable $\mathbf{b}$ & \bmcell{48.5}{3.7} \\
Disable $\mathbf{W}$ &\bmcell{41.2}{1.8} \\
Scaling channels only & \bmcell{43.3}{3.2} \\
\bottomrule
\end{tabular}
\end{table}

\begin{figure}[h!]
    \centering
    \includegraphics[width=1\linewidth]{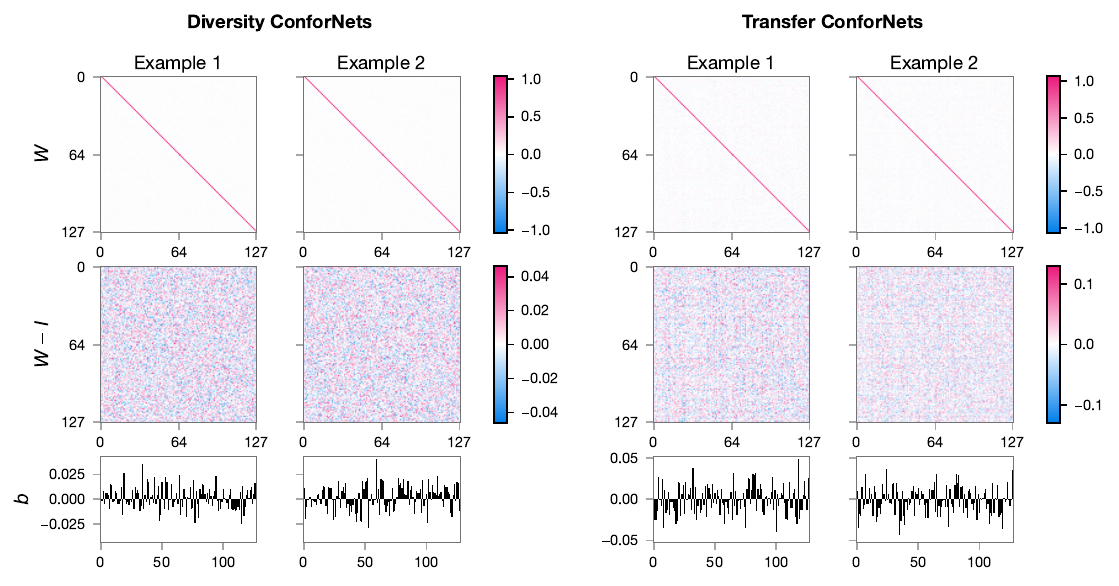}
    \caption{\textbf{Representative examples of learned ConforNets for centroid protein of membrane transporters benchmark}, uncharacterized ABC transporter ATP-binding protein TM\_0288. Each column shows one 128-dimensional ConforNet $(\mathbf{W},\mathbf{b})$; the first two columns are examples drawn from diversity-maximizing training (20 steps), and the last two are examples from transfer training. Rows visualize, from top to bottom: the learned linear weight $\mathbf{W}$, its deviation from the identity $\mathbf{W} - \mathbf{I}$, and the bias vector $\mathbf{b}$.}
    \label{fig:app_wb_examples}
\end{figure}

\clearpage 
Figure~\ref{fig:app_wb_examples} visualizes representative learned ConforNets from both diversity and transfer settings.

\begin{figure}[h!]
    \centering
    \includegraphics[width=0.7\linewidth]{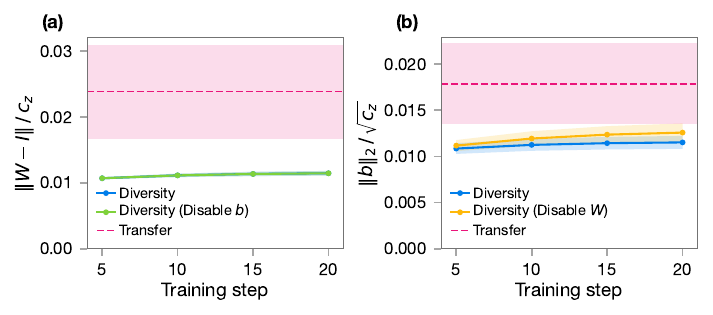}
    \caption{\textbf{Magnitude of learned ConforNet parameters $(W,b)$ across training steps.} The solid lines show the mean across diversity ConforNets, and shaded bands indicate standard deviation across diversity ConforNets at each training step. The dashed horizontal line marks the mean of the per-ConforNet metric across all transfer runs, and the pink horizontal band shows standard deviation across transfer ConforNets. \textbf{(a)} L2 norm of the weight deviation from identity $\mathbf{W}-\mathbf{I}$, normalized by $c_z$. \textbf{(b)} L2 norm of the bias $b$, normalized by $\sqrt{c_z}$.}
    \label{fig:per_step_wb_vs_transfer}
\end{figure}

Figure~\ref{fig:per_step_wb_vs_transfer} shows that ConforNet magnitudes increase only modestly over the 20 diversity training steps considered here and remain substantially smaller than those of transfer ConforNets, suggesting that the gap may reflect both longer training and differences in learning signal.

\clearpage
\section{Interpreting Pairformer latents}
\label{app:foldswitch_interp}

\begin{figure}[h!]
    \centering
    \includegraphics[width=0.75\linewidth]{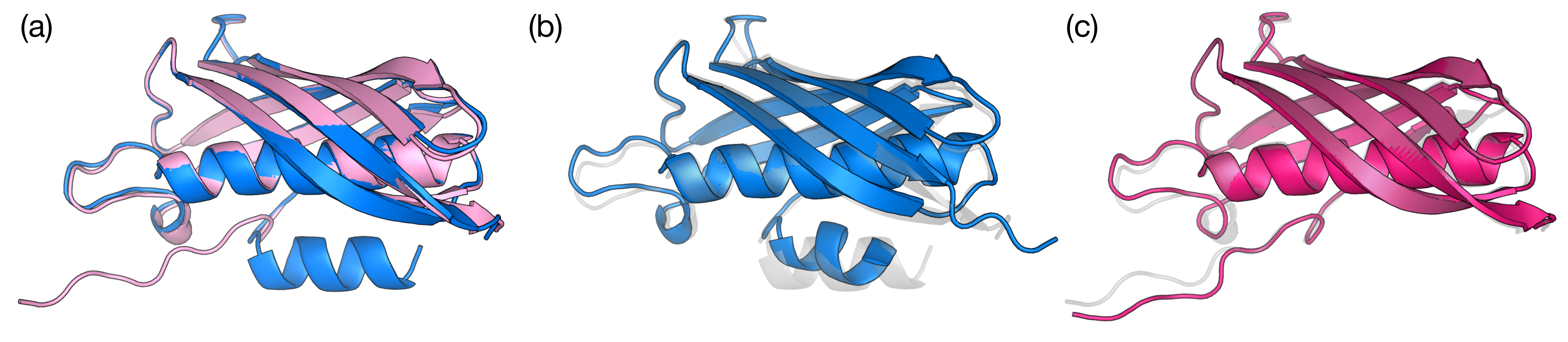}
    \caption{\textbf{(a)} Ground-truth N-terminal helix (\href{https://www.rcsb.org/structure/4ZRB}{4ZRB};H) and N-terminal coil (\href{https://www.rcsb.org/structure/4ZRB}{4ZRB};C) conformations of PaaI Thioesterase superimposed. \textbf{(b)} Structure induced by one ConforNet, which is RMSD = 1.38\AA{} to 4ZRB;H (overlaid). \textbf{(c)} Structure induced by the other ConforNet, which is RMSD = 1.29\AA{} to 4ZRB;C (overlaid).}
    \label{fig:interp_gts}
\end{figure}

In this section, we examine how ConforNets applied to the pre-Pairformer pair representation propagate their effects through Pairformer blocks. As a case study, we focus on the fold-switching protein PaaI thioesterase, whose two known conformations differ by a rearrangement of the N-terminal helix (Fig.~\ref{fig:interp_gts}a). This conformational change repositions the Tyr38/Tyr39 loop, which gates substrate access \citep{khandokar2016structural}. We selected this example because the conformational difference is localized and easily structurally interpretable. 

We obtained two ConforNets, each producing samples close to one of the two reference conformations, using the coordinate MSE objective (unsupervised). As shown in Fig.~\ref{fig:interp_gts}, the induced structures closely recapitulate the conformational switch.

\begin{figure}[t]
    \centering
    \begin{subfigure}[t]{0.3\linewidth}
        \centering
        \includegraphics[width=\linewidth]{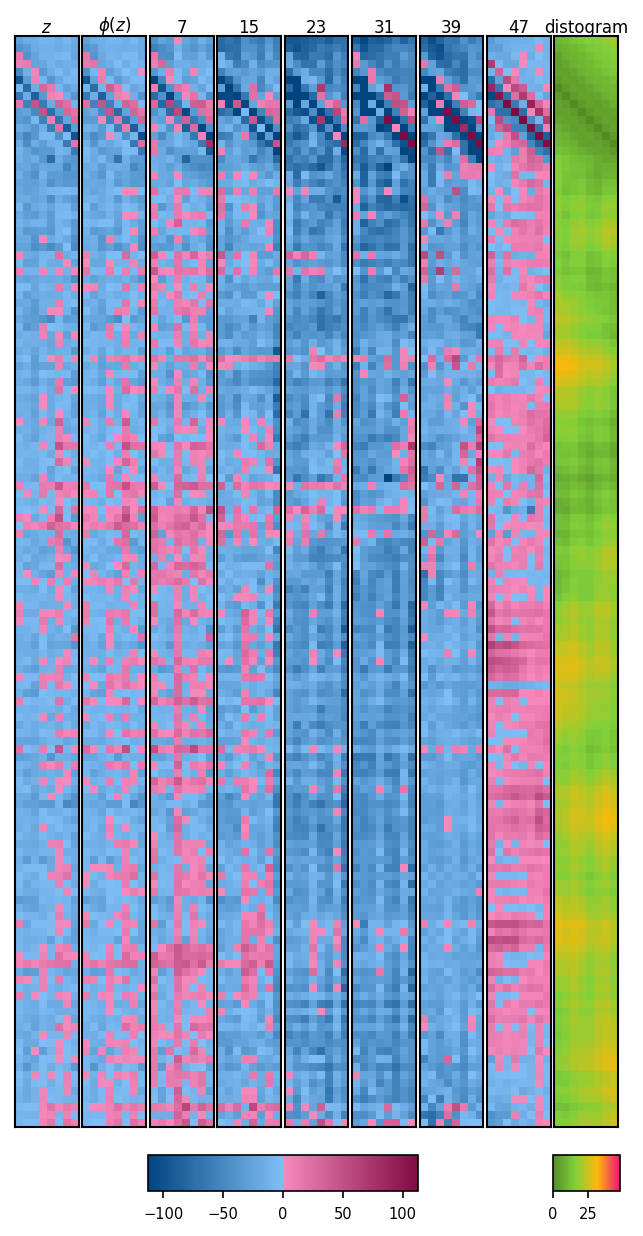}
        \caption{Evolution under $\phi_H$}
    \end{subfigure}
    \hfill
    \begin{subfigure}[t]{0.3\linewidth}
        \centering
        \includegraphics[width=\linewidth]{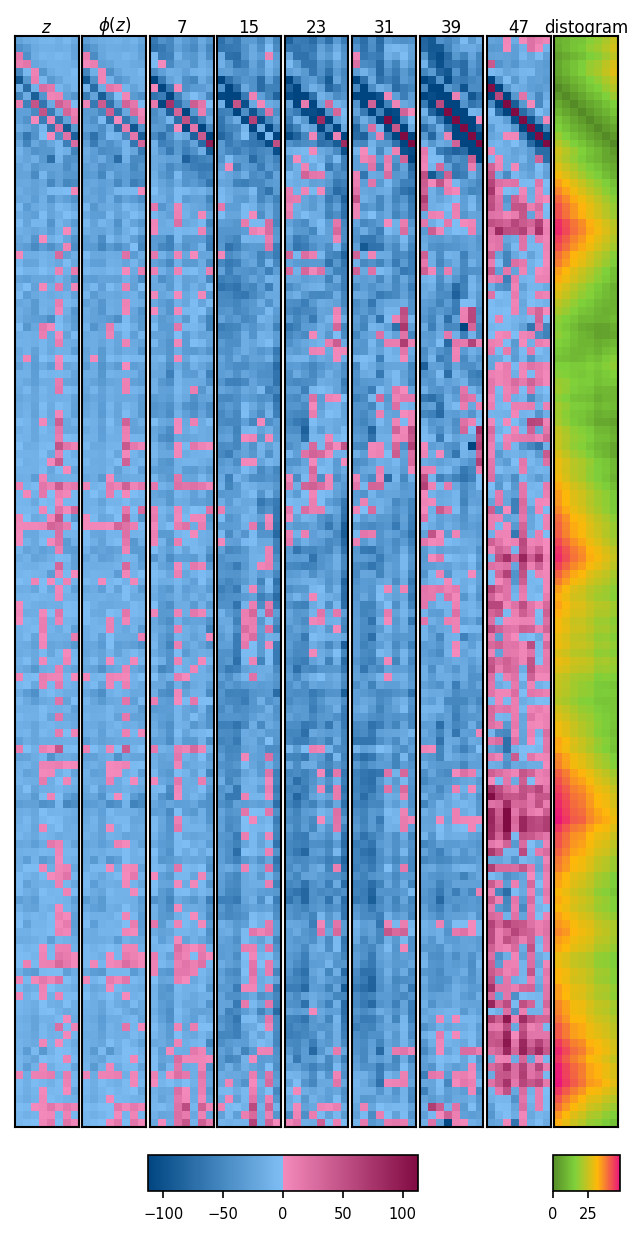}
        \caption{Evolution under $\phi_C$}
    \end{subfigure}
    \hfill
    \begin{subfigure}[t]{0.3\linewidth}
        \centering
        \includegraphics[width=\linewidth]{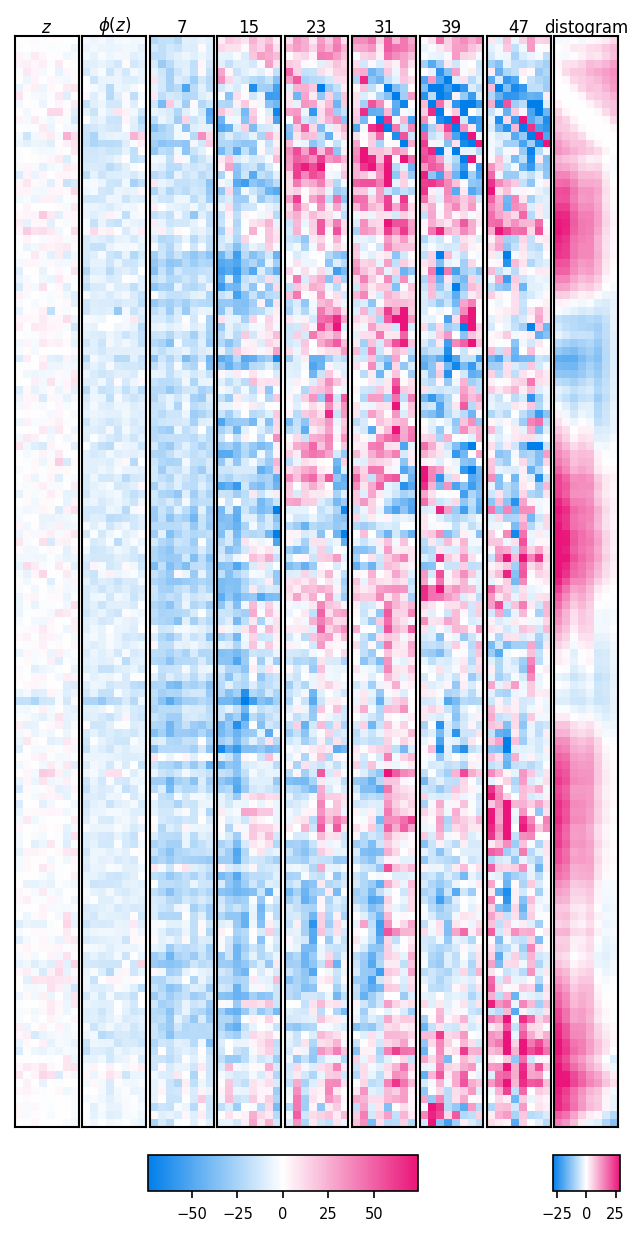}
        \caption{Difference}
    \end{subfigure}
    \caption{Evolution of the pair representation slice in \textbf{channel 0} under $\phi_H$ and $\phi_C$, and their difference, across Pairformer depth.}
    \label{fig:viz_channel_0}
\end{figure}

\begin{figure}[t]
    \centering
    \begin{subfigure}[t]{0.3\linewidth}
        \centering
        \includegraphics[width=\linewidth]{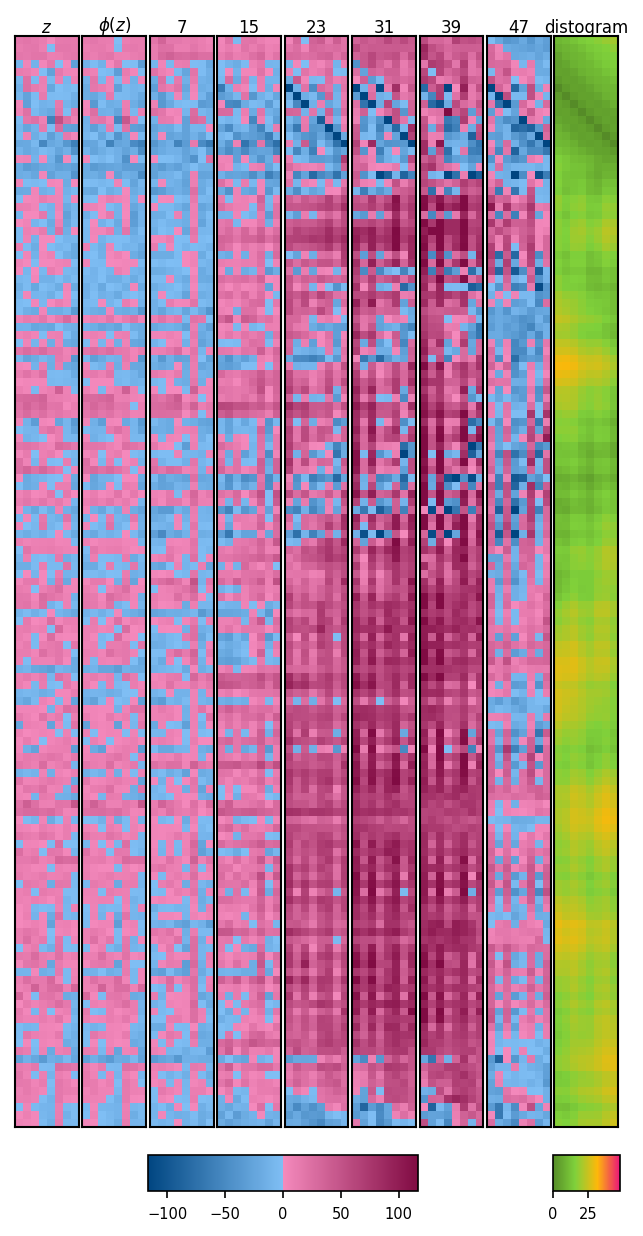}
        \caption{Evolution under $\phi_H$}
    \end{subfigure}
    \hfill
    \begin{subfigure}[t]{0.3\linewidth}
        \centering
        \includegraphics[width=\linewidth]{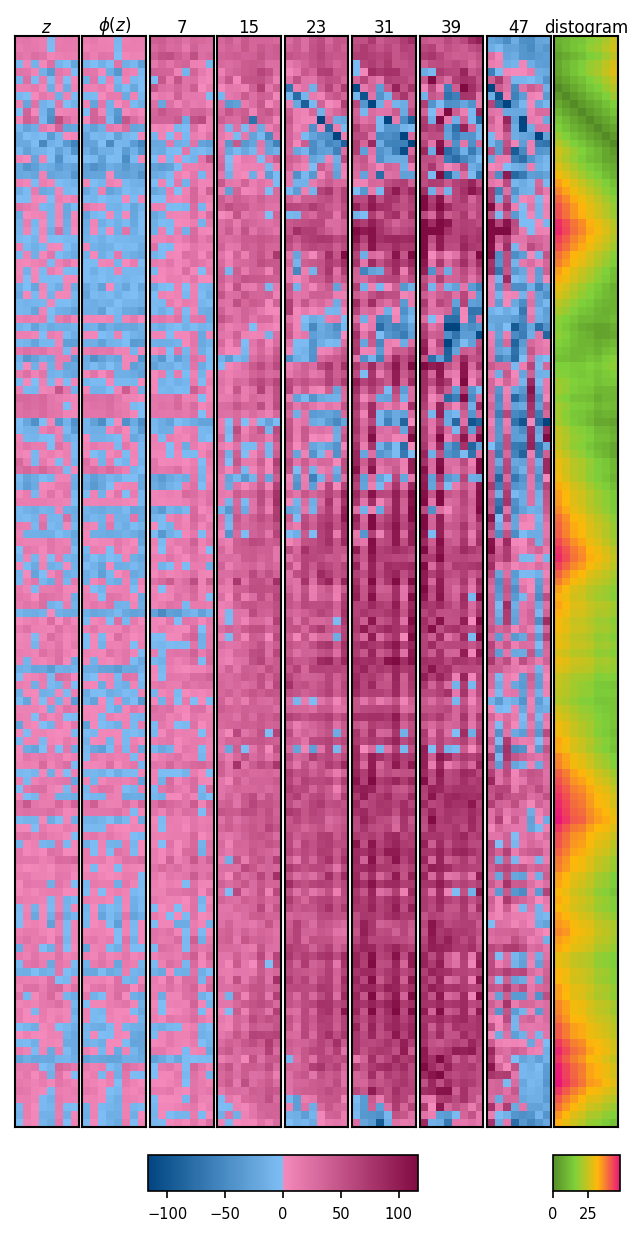}
        \caption{Evolution under $\phi_C$}
    \end{subfigure}
    \hfill
    \begin{subfigure}[t]{0.3\linewidth}
        \centering
        \includegraphics[width=\linewidth]{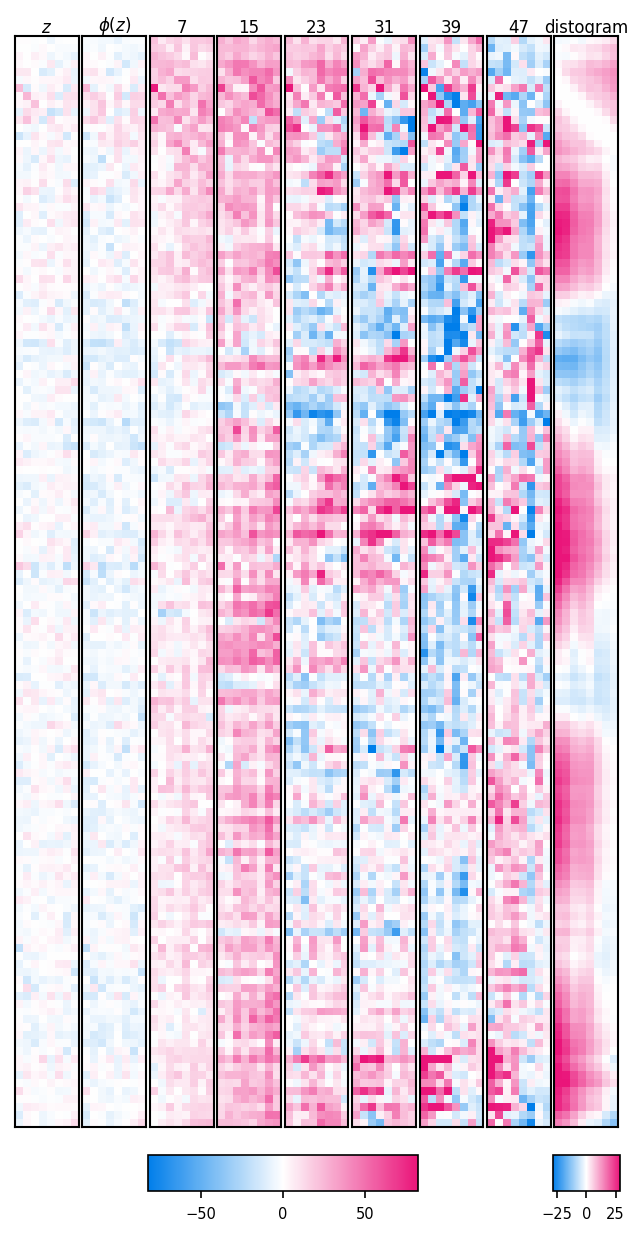}
        \caption{Difference}
    \end{subfigure}
    \caption{Evolution of the pair representation slice in \textbf{channel 1} under $\phi_H$ and $\phi_C$, and their difference, across Pairformer depth.}
    \label{fig:viz_channel_1}
\end{figure}

We denote the two learned ConforNets $\phi_H$ and $\phi_C$, as they induce predictions close to the H and C conformations, respectively. To visualize how the pair representation evolves to produce this conformational change, we focus on an $8 \times 137$ slice formed by the N-terminal helix that rearranges against all 137 residues in the protein. Figures~\ref{fig:viz_channel_0} and \ref{fig:viz_channel_1} show this slice for channels 0 and 1, respectively. In each figure, panels (a) and (b) show the evolution of the slice under $\phi_H$ and $\phi_C$, while panel (c) shows their difference. We show the slice before adaptation ($\mathbf{z}^{\mathrm{pre}}$), after adaptation ($\phi(\mathbf{z}^{\mathrm{pre}})$), and after selected Pairformer blocks (0-indexed). The rightmost column of each panel shows the ground truth distogram slice: 4ZRB;H in (a), 4ZRB;C in (b), and their difference in (c).

The pattern resembling the ground truth distogram difference is not clearly visible in the early Pairformer blocks. Instead, it gradually emerges and sharpens with depth, suggesting that the ConforNet does not directly write the final contact pattern into $\mathbf{z}^{\mathrm{pre}}$, but rather biases the Pairformer toward evolving distinct conformational features over successive blocks.


\end{document}